\documentclass[12pt]{article}
\usepackage[version=3]{mhchem} 
\usepackage{graphicx,amsmath,amssymb,epsfig}
\usepackage{gensymb} 
\usepackage[cp1252]{inputenc}
\usepackage[super,sort&compress,comma]{natbib}

\usepackage{times}

\usepackage[]{units}

\title{Configuration of the magnetosome chain: a natural magnetic nanoarchitecture}

\author{I. Orue,$^{1}$ L. Marcano,$^{2}$ P. Bender,$^{3}$ A. Garc\'{\i}a-Prieto,$^{4,5}$ S. Valencia,$^{6}$\\ M.A. Mawass,$^{6}$ D. Gil-Cart\'{o}n,$^{7}$ D. Alba Venero,$^{8}$ D. Honecker,$^{9}$\\ A. Garc\'{\i}a-Arribas,$^{2,5}$ L. Fern\'{a}ndez Barqu\'{\i}n,$^{3}$\\ A. Muela,$^{5,10}$ M.L. Fdez-Gubieda$^{2,5\ast}$\\
\\
\normalsize{$^{1}$SGIker, Universidad del Pa\'{\i}s Vasco - UPV/EHU, 48940 Leioa, Spain,}\\
\normalsize{$^{2}$Dpto. Electricidad y Electr\'{o}nica, Universidad del Pa\'{\i}s Vasco - UPV/EHU, 48940 Leioa, Spain}\\
\normalsize{$^{3}$CITIMAC, Universidad de Cantabria, 39005 Santander, Spain}\\
\normalsize{$^{4}$Dpto. F\'{\i}sica Aplicada I, Universidad del Pa\'{\i}s Vasco - UPV/EHU, 48013 Bilbao, Spain}\\
\normalsize{$^{5}$BCMaterials, Bld. Martina Casiano, 48940 Leioa, Spain}\\
\normalsize{$^{6}$Helmholtz-Zentrum Berlin f\"{u}r Materialien und Energie,}\\
\normalsize{Albert-Einstein-str. 15, 12489 Berlin, Germany}\\
\normalsize{$^{7}$Structural Biology Unit, CIC bioGUNE, CIBERehd, 48160 Derio, Spain}\\
\normalsize{$^{8}$ISIS, STFC Rutherford Appleton Laboratory, Chilton,}\\
\normalsize{Didcot OX11 0QX, United Kingdom}\\
\normalsize{$^{9}$Institut Laue-Langevin, 38042 Grenoble, France}\\
\normalsize{$^{10}$Dpto. Inmunolog\'{\i}a, Microbiolog\'{\i}a y Parasitolog\'{\i}a,}\\
\normalsize{Universidad del Pa\'{\i}s Vasco - UPV/EHU, 48940 Leioa, Spain}\\
\\
\normalsize{$^\ast$To whom correspondence should be addressed; e-mail:  malu.gubieda@ehu.eus.}
}

\begin{document}

\baselineskip24pt

\maketitle


\newpage

\begin{abstract}
\normalsize{\emph{Magnetospirillum gryphiswaldense} is a microorganism with the ability to bio\-mi\-ne\-ra\-lize magnetite nanoparticles, called magnetosomes, and arrange them into a chain that behaves like a magnetic compass. Rather than straight lines, magnetosome chains are slightly bent, as evidenced by electron cryotomography. Our experimental and theoretical results suggest that due to the competition between the magnetocrystalline and shape anisotropies, the effective magnetic moment of individual magnetosomes is tilted out of the [111] crystallographic easy axis of magnetite. This tilt does not affect the direction of the chain net magnetic moment, which remains along the [111] axis, but explains the arrangement of magnetosomes in helical-like shaped chains. Indeed, we demonstrate that the chain shape can be reproduced by considering an interplay between the magnetic dipolar interactions between magnetosomes, ruled by the orientation of the magnetosome magnetic moment, and a lipid/protein-based mechanism, modeled as an elastic recovery force exerted on the magnetosomes.
}
\end{abstract}

Magnetotactic bacteria (MTB) are microorganisms ubiquitously present in any aquatic environment with the capability of magnetic orientation~\cite{Blakemore1975}. The origin of this behavior is  the presence of intracellular magnetic nanoparticles enclosed in a vesicle, called magnetosomes. The size, morphology and  composition of the magnetosomes depend on the species, being the size, in all of them, large enough to be magnetically stable at room temperature~\cite{DuninBorkowski1998}. Most MTB organize the magnetosomes in a chain. Since magnetosomes are single magnetic domains, the chain behaves as a large single permanent dipole magnet which is able to passively reorient the whole bacteria along external magnetic field lines, even at fields as small as those existing near the Earth's surface ($\approx \unit[20-60]{\mu T}$)~\cite{Bazylinski2004,Uebe2016,Lefevre2013}. On top of the fundamental scientific interest of these fascinating microorganisms, magnetosome chains constitute a natural paradigm of highly anisotropic magnetic nanostructures, which show high potentiality in biomedical applications~\cite{Serantes2014,Felfoul2016,Ghosh2012,Alphandery2011}, in actuation devices as nanorobots~\cite{Mishra2016}, in nanosensor devices~\cite{Jiang2016}, and in magnetic memory devices~\cite{Kou2011}.

Among all the MTB species, \emph{Magnetospirillum gryphiswaldense} strain MSR-1 is frequently used as a model strain of MTB  due to its ease of cultivation in the laboratory. \emph{Magnetospirillum} cells are helical and contain a variable number of $\sim \unit[40-45]{nm}$ sized cuboctahedral magnetite (\ce{Fe3O4}) magnetosomes arranged in a chain. The biomineralization of the magnetite nanoparticles and the assembly in a chain are complex and comprise different steps~\cite{Uebe2016,Komeili2007,Faivre2008}. Firstly, the magnetosome membrane is formed from invagination of the cytoplasmic membrane comprised with proteins~\cite{Komeili2006}. Secondly, iron is taken up from the environment and transported into the magnetosome vesicle where nucleation in magnetite nanocrystals occurs~\cite{Frankel1983,FdezGubieda2013,Baumgartner2013}, and  finally, magnetosomes  are assembled forming the chain~\cite{Uebe2016,Komeili2012,Toro2016,Cornejo2016}. The present work is devoted to shed light on this last step. In particular, here we address the underlying mechanisms that determine the arrangement of the magnetosomes and consequently the geometry of the chain.

The arrangement of magnetosomes in a chain results from the interplay between the active assembly mechanism mediated by proteins and the magnetic dipolar interactions between nearest magnetite nanoparticles~\cite{Shcherbakov1997,Klumpp2012,Meyra2016}.
As shown by cryotomography imaging on \emph{Magnetospirilla}, bundles of cytoskeletal filaments intervene in the chain assembly~\cite{Komeili2006,Katzmann2010,Scheffel2006}. These filaments, formed by the actin-like protein MamK, traverse the cell and position the chain in the middle of the cell. Another important protein involved in the chain formation is MamJ. One possible function of MamJ is to connect the magnetosome membrane to the cytoskeletal filament. As magnetosomes get closer together, magnetic dipolar interactions arise. 

Here we show by electron cryotomography imaging that magnetosome chains of \emph{M. gry\-phis\-wal\-den\-se} are not straight lines but appear slightly bent, a fact that has been also observed in
previous works~\cite{Faivre2008,Scheffel2006}. The implications of the chain shape on the magnetic response of magnetosome chains have then been addressed with complementary techniques performed on a set of bacterial arrangements: i) small angle neutron/x-ray scattering (SANS/SAXS) on a bacterial colloid, ii) macroscopic magnetometry on 3D and 2D fixed arrangements of randomly distributed and aligned bacteria, and iii) x-ray photoemission electron microscopy (XPEEM) on an individual chain of magnetosomes extracted from bacteria.

Two main findings are achieved from this work. Firstly, the  equilibrium magnetic moment of the magnetosomes is tilted 20\degree\ out of the [111] crystallographic axis, as concluded from the magnetic analysis. Secondly, the tilting of the magnetic moment is the key to understand the helical-like shape of the magnetosome chains. Indeed, the experimental chain patterns imaged by  electron cryotomography are accurately reproduced  by counterbalancing the magnetic dipolar interactions between magnetosomes, strongly affected by the orientation of the individual magnetic moments, and
a lipid/protein-based mechanism, modeled as an elastic recovery force exerted on magnetosomes.

Besides the basic interest of this study, a precise knowledge of the mechanisms determining the chain shape is decisive in applications of MTB such as biological micro-robots. Precisely, recent works propose MTB to be exploited as motors with embedded source propulsion secured by their flagella and embedded steering system to control the directional motion provided by their magnetosome chain~\cite{Felfoul2016}.

\section*{Results and discussion}

\subsection*{Electron cryotomography imaging of the magnetosome chain}\label{ECT}

\begin{figure}
\centering
\includegraphics[scale=0.7]{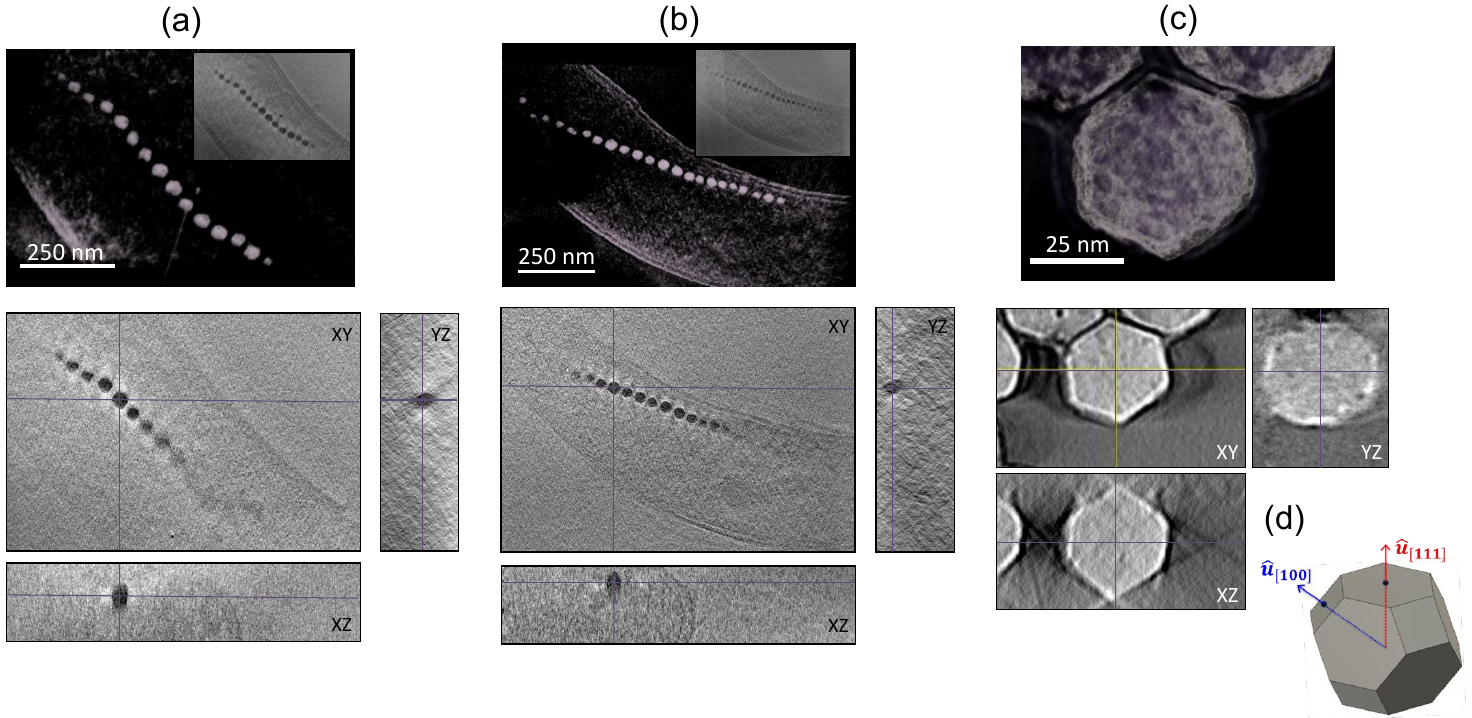}\\
\caption{Electron cryotomography (ECT) of magnetosome chains. (a) and (b) are ECT images of two different bacteria.  Top: 3D tomograms reconstructed from the tilt-series of images obtained by ECT. Videos of the tomograms of the two chains viewed from different perspectives (as consecutive slices in XY, XZ and YZ plane orientation, and as an electron density map) can be found in the supplementary information (movies S1-S2). The insets show the Z-projection of the tomograms. Bottom: XY, YZ and XZ slices of the tomograms shown on top. The three slices are taken along the marked lines and cross on one of the magnetosomes.
c) ECT images of extracted magnetosomes. Top: reconstructed tomograms. Bottom: central XY, YZ and XZ slices of the magnetosome tomogram shown on top.
d) Schematic representation of a cuboctahedral magnetosome.}
\label{fig:cryotomo2}\end{figure}

Cells of \emph{M. gryphiswaldense} have been imaged by electron cryotomography (ECT). Since bacteria are cryoembedded, the cells and as a consequence the magnetosome chains inside them preserve their natural shape, avoiding artifacts associated to cell drying processes. ECT images of two cells are shown in Figs.~\ref{fig:cryotomo2}a,b.
The upper part of Fig.~\ref{fig:cryotomo2}a,b shows the reconstructed 3D tomograms from the tilt-series of images obtained by ECT of the two cells and
the insets show the $Z$-projected images of the tomograms. These projections are equivalent to the images obtained by transmission electron microscopy. The chains shown in Fig.~\ref{fig:cryotomo2} are composed of $N=15$ and $22$ magnetosomes, respectively, and the mean distance between magnetosomes is $d \approx \unit[60]{nm}$. Videos showing the 3D reconstructions of the two magnetosome chains from different perspectives can be found in the supplementary information (movies S1-S2). ECT imaging evidences that magnetosome chains are certainly not straight lines but exhibit more complex  patterns. Three slices (XY, YZ and XZ) of each of the 3D reconstructions are shown below the tomograms in Fig.~\ref{fig:cryotomo2}a,b. The three slices are cuts of the corresponding tomograms taken along the marked lines and intersect in one of the magnetosomes, so that the XYZ coordinates of that particular magnetosome can be determined. By determining the XYZ coordinates of all the magnetosomes in the chain we have been able to reconstruct the two magnetosome chains, which are shown in Fig.~\ref{fig:chain_config}b, where a zoom-in of one of the magnetosome chain reconstructions emphasizes the deviation from a straight line.

ECT images of a magnetosome extracted from \emph{M. gryphiswaldense} reveal the morphology of the magnetosome (Fig.~\ref{fig:cryotomo2}c). The figure on the top is a reconstructed 3D tomogram, and below, the XY, YZ and XZ central sections of the tomogram are shown. These images evidence that magnetosomes have a strongly faceted morphology similar to the truncated octahedral one, which is adopted as a reference (Fig.~\ref{fig:cryotomo2}d), with a mean size of $\unit[40]{nm}$. In this truncated octahedron the $\langle 001 \rangle$ crystallographic axes define the growth directions of the square faces and the $\langle 111 \rangle$ those of the hexagonal faces. The magnetosome in the image has two neighbours (only partially observed in the image). The three magnetosomes have self-assembled so that their hexagonal faces are towards each other. In the same way, when magnetosomes are forming the chain inside the cell, numerous studies in \emph{Magnetospirilla} show that they align with their hexagonal faces towards each other~\cite{Mann1984,Kornig2014}, i.e., along one of their $\langle 111 \rangle$ crystallographic directions, as sketched in Fig.~\ref{fig:chain_config}a,d. The [111] direction along which magnetosomes align defines the so-called \emph{chain axis}.

\subsection*{Magnetism of magnetosome chains}

ECT images of the magnetosome chain show a complex  structure and an evident deviation from a straight linear configuration.  This must have an effect on the magnetic behaviour of the chain. We have analyzed
the magnetic properties of magnetosome chains from \emph{M. gryphiswaldense}  by complementary magnetic techniques in three different arrangements: i) a colloidal dispersion of bacteria, this means that we have a 3D distribution of bacterial chains, studied by magnetization and small angle neutron/x-ray scattering (SANS/SAXS); ii) macroscopic magnetization measurements on bacteria forming 3D and 2D distributions; iii) x-ray photoemission electron microscopy (XPEEM) on individual chains of magnetosomes extracted from the bacteria forming a 1D magnetosome arrangement.

\subsubsection*{Magnetization and small angle neutron/x-ray scattering (SANS/SAXS) on a bacterial colloid}
\label{sans}

\begin{figure}
\centering
\includegraphics[clip,scale=0.7]{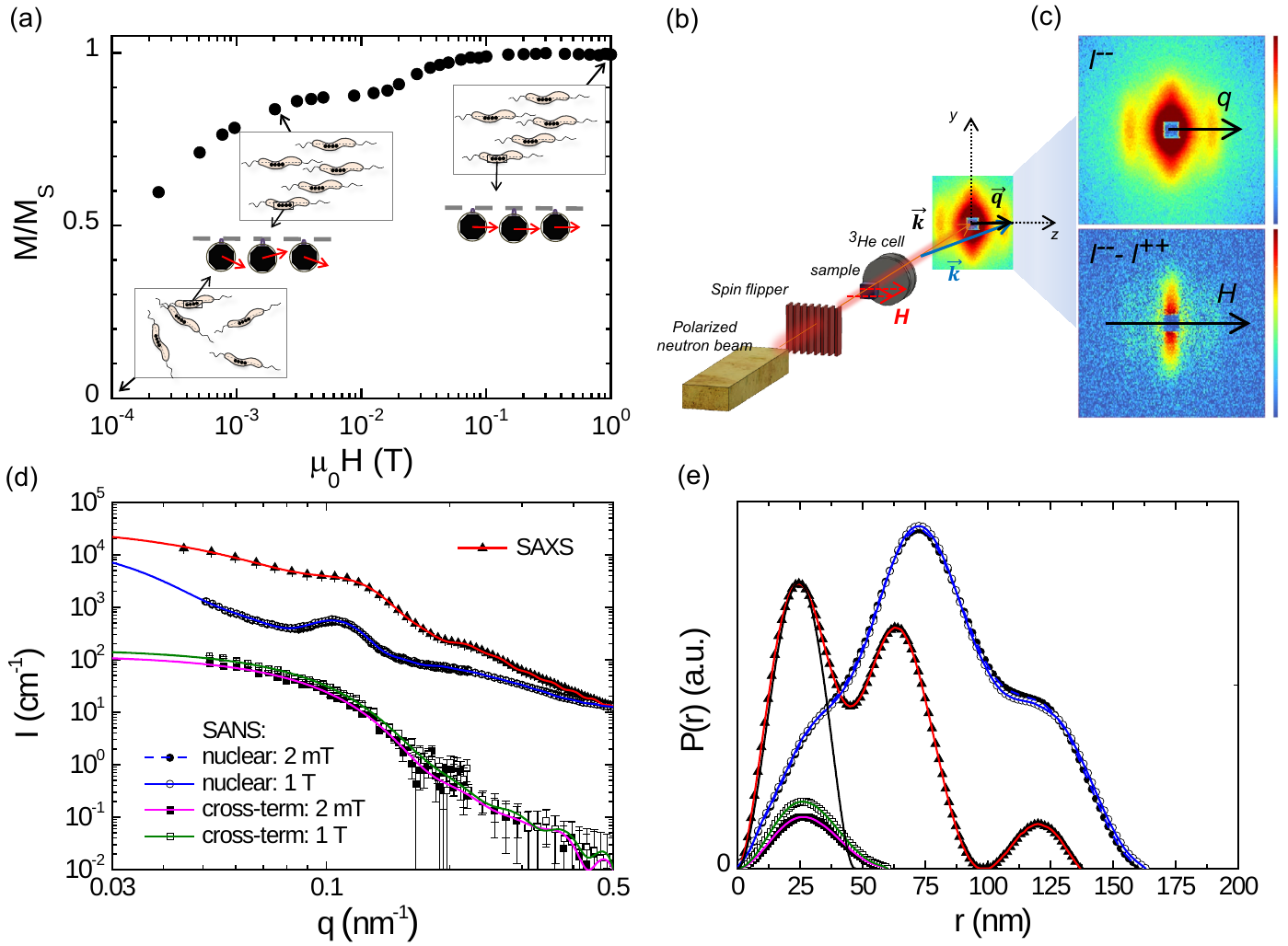}\\
\caption{Magnetic state of the colloidal dispersion of bacteria.
a) Magnetization curve of the colloid. The field axis is in logarithmic scale to magnify the low-field region. The sketches display the two-step magnetization process. The experimental data marked with arrows correspond to the points measured by SANS/SAXS ($\mu_0 H=\unit[0]{mT}; \unit[2]{mT}$; and $\unit[1]{T}$).
b) Schematic representation of the SANS experiment. The polarized incoming neutron beam can be set to either parallel (+) or antiparallel (-) to the applied field by means of an RF spin flipper. The $^3$He cell discriminates the polarization of the scattered neutrons (+ or -), hence the recorded intensity is either $I^{++}$ or $I^{--}$, where superindexes refer to the polarization of the incoming/scattered neutrons.
c) 2D SANS scattering patterns for $\mu_0 H=\unit[2]{mT}$. Top: $I^{--}$; bottom: $I^{--}-I^{++}$.
d) 1D scattering intensities measured by SAXS in zero field (radial average, offset by scale factor 100) and the field dependent nuclear scattering intensities $I_{nuc}(q)$ (offset by scale factor 10) as well as the cross-terms $I_{cross}(q)$ determined by polarized SANS.
The lines are the corresponding fits by an indirect Fourier transform.
e) Pair distance distribution functions $P(r)$ determined by an indirect Fourier transform of the 1D scattering intensities from (d).
The black line is the $P(r)$ of a homogeneous sphere with diameter $D_{SANS}=\unit[48]{nm}$.
The distribution function determined by SAXS is offset by arbitrary scaling factors.
}\label{fig:SANS}
\end{figure}

Fig.~\ref{fig:SANS}a shows the magnetization curve $M(H)$ of a colloidal dispersion of bacteria ($\unit[6\cdot10^{11}]{cell/mL}$). This curve was measured slowly, leaving two minutes between each data point to assure thermal equilibrium is achieved at each point. As shown in Fig.~\ref{fig:SANS}a, the magnetization increases rapidly with the applied field until it reaches a plateau between $\unit[3]{mT} \le \mu_0H \le \unit[15]{mT}$ at a value that is 90\% of the saturation value ($M=0.90 M_s$). Then the magnetization increases again slowly up to saturation. This behavior reveals that there are two different coherent rotations of the magnetization as will be discussed in the following.

The magnetic properties of the bacterial colloid have been further investigated via small angle x-ray scattering (SAXS) and polarized small angle neutron scattering (SANS). In a SAXS/\-SANS experiment the x-ray/neutron beam ($\vec{k}_{IN}$) hits the sample and the scattered beam ($\vec{k}_{scat}$) impacts on a 2D detector perpendicular to the incoming beam (Fig.~\ref{fig:SANS}b). The intensity $I$ as a function of the scattering vector $\vec{q}=\vec{k}_{scat}-\vec{k}_{IN}$ gives pure structural ($I_{nuc}(q)$) information from SAXS/SANS experiments, or, additionally, magnetic ($I_{cross}(q)$) information for SANS depending on the magnetic field on the sample and the polarization of the incoming and scattered neutron beam (Fig.~\ref{fig:SANS}c). $I_{nuc}(q)$ and $I_{cross}(q)$ are extracted from the 2D scattering patterns (Fig.~\ref{fig:SANS}c) as explained in the methods section.

Fig.~\ref{fig:SANS}d shows the radially averaged 1D SAXS intensity $I(q)$ of the colloidal dispersion measured in zero field.
An indirect Fourier transform of $I(q)$ results in the pair distance distribution function $P(r)$ displayed in Fig.~\ref{fig:SANS}e.
The extracted distribution function exhibits three distinct peaks.
The first peak has its maximum at about \unit[25]{nm} and is nearly bell-shaped.
The comparison of this peak with the pair distance distribution function of a homogeneous sphere with diameter $D_{SANS}=\unit[48]{nm}$ shows qualitatively very good agreement.
Hence, we surmise that the first peak corresponds to the nuclear scattering of the individual magnetosome of an average size of about $\unit[48]{nm}$. This particle size is considerably larger than the one obtained by ECT ($\approx \unit[40]{nm}$). The reason for this difference is that ECT is only sensitive to the core of the magnetosome (the magnetic nanoparticle), while SANS is sensitive to both the core and the surrounding lipid bilayer membrane. In fact, for neutrons, the scattering cross section of H is very large, which means that organic materials with many H-atoms (such as the lipid membrane in our case) can generate large scattering signals \cite{Hoell2004}. Considering a membrane thickness of $\approx \unit[4]{nm}$, the core diameter obtained by SANS is $\unit[48-8=40]{nm}$, in agreement with the value obtained from ECT.
The maxima of the second and third peak of the distribution function are at about $75$ and $\unit[125]{nm}$, and both peaks are also nearly bell-shaped.
The positions of the peaks agree well with the expected positions of the center of mass of the next neighbors in a chain-like structure of $D_{SANS} = \unit[48]{nm}$ sized particles with a center-to-center distance of $d \approx \unit[50]{nm}$, close to  the $d \approx \unit[60]{nm}$ distance measured by ECT.

Fig.~\ref{fig:SANS}d additionally shows the structural 1D SANS intensities $I_{nuc}(q)$ detected for field strengths of $\mu_0H=\unit[2]{mT}$ and $\unit[1]{T}$ applied perpendicularly to the neutron beam. 
Both $I_{nuc}(q)$'s are virtually identical, which means that already at \unit[2]{mT} the bacteria are fully aligned in the field direction.
Consequently an indirect Fourier transform of both nuclear scattering intensities results in superimposed pair distance distribution functions, whose second peak ($r \sim \unit[75]{nm}$) is about twice the height of the first peak (Fig.~\ref{fig:SANS}e).
This verifies that the bacteria align with the chain axis parallel to the field, considering that within the chain each particle is surrounded on average by two neighbors and thus the probability to find a scatterer at this position is two times the probability to find a scatterer within the primary particle (first peak).
To investigate if structural alignment equals magnetic saturation the cross-terms $I_{cross}(q)$ were analyzed.

The two cross-terms $I_{cross}(q)$ detected at \unit[2]{mT} and \unit[1]{T} display the same functional form (Fig.~\ref{fig:SANS}d) and accordingly the extracted distribution functions are qualitatively similar (Fig.~\ref{fig:SANS}e).
The observation of only one peak can be explained by the fact that the cross terms depict the correlation between nuclear scattering length density and magnetization $M_z(r)$ along the axis perpendicular to the applied field. 
In all cases the shape of the distribution function is comparable to the distribution function of a single sphere with a homogeneous scattering length density.
This verifies that the particles are homogeneously magnetized (i.e. $M_z(r)=M_z$) and thus can be regarded as single-domain particles.
However, the absolute values of $P(r)$ detected at $\unit[2]{mT}$ are over the whole $r$-range systematically reduced by a factor of $0.83$ compared to $\unit[1]{T}$.
Assuming that at $\mu_0H=\unit[1]{T}$ the system is saturated in field direction (i.e. $M_z=M_s$) this means that at $\unit[2]{mT}$ the magnetization in field direction amounts to $M_z=0.83 M_s$ and that the magnetic saturation is achieved by a coherent rotation of the spins within the individual nanoparticles towards the field direction. These two processes (chain alignment followed by a coherent rotation of the  magnetosome spins) explain the isothermal magnetization measurement of the colloid shown in Fig.~\ref{fig:SANS}a. In the latter, in agreement with the SANS result, at $\unit[2]{mT}$ $M_z=0.83 M_s$, and the coherent rotation of magnetosome spins would start at $\approx \unit[15]{mT}$, where $M_z=0.9 M_s$ corresponds to a misalignment of the magnetic moment with the chain axis of $\theta \approx 25 \degree$, where $M_z=M_s \cos \theta$.

\subsubsection*{Magnetometry on 2D and 3D bacterial arrangements}
\label{squid}

Macroscopic hysteresis loops of \emph{M. gryphiswaldense} cells have been measured by SQUID and VSM magnetometry. 3D arrangements of aligned bacteria have been obtained by pouring the cells under an applied 'aligning' uniform magnetic field ($\vec{H}_{al}$) into liquid agar that hardens upon cooling.  Similarly, 2D arrangements of aligned bacteria have been obtained by depositing the cells onto a Si substrate under $\vec{H}_{al}$. A TEM image of aligned \emph{M. gryphiswaldense} cells deposited onto a Si substrate is shown in the supplementary information.
Samples of randomly arranged bacteria have been also prepared in both the 3D and 2D configurations.

\begin{figure}
\centering
\includegraphics[clip,scale=1]{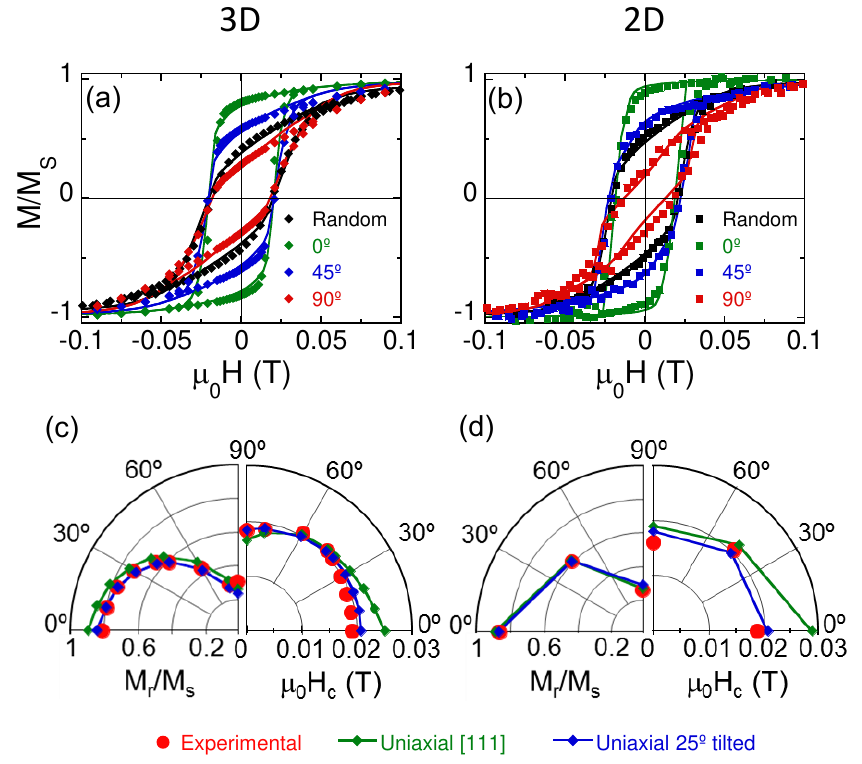}\\
\caption{
 Hysteresis loops of oriented bacteria under an applied field in 3D (a) and 2D (b) arrangements  forming 0\degree, 45\degree, and 90\degree\ with the aligning field ($\vec{H}_{al}$), together with the hysteresis loops of randomly arranged bacteria. The continuous line is the fit to the magnetic model explained in the text.
c) and d) Polar plots of the coercivity and reduced remanent magnetization at different orientation angles of the chain  axis with $\vec{H}_{al}$ in the 3D (c) and 2D (d) arrangements. The experimental points are compared to the results of two simulations in which the uniaxial easy axis $\hat{u}_{uni}$ is either parallel (green) or tilted 25\degree\ (blue) with the chain  axis.
}\label{fig:SQUID}
\end{figure}

The hysteresis loops of randomly arranged and oriented bacteria are shown in Fig.~\ref{fig:SQUID}a,b. The hysteresis loops of oriented bacteria have been measured at different angles with respect to $\vec{H}_{al}$ between 0\degree and 90\degree in steps of 20 degrees for 3D arrangement and at discrete angles, 0\degree, 45\degree\ and 90\degree, for 2D arrangement. These measurements evidence that bacteria are highly anisotropic magnetic objects, since their hysteresis loops depend strongly on the direction of the applied field. This is not surprising, as one would expect that magnetosome chains behaved as a good compass, with a single magnetic easy axis oriented along the chain axis,  which, as noted previously, is coincident with one of the magnetosomes $\langle 111 \rangle$  crystallographic axes. However, as previously observed in \emph{Magnetospirillum magneticum} AMB-1~\cite{Li2013}, the hysteresis loops of 3D and 2D bacterial arrangements do not correspond to those expected for a uniaxial Stoner-Wohlfarth model~\cite{Stoner1948} along the chain, since the hysteresis loop perpendicular to the chain axis is not anhysteretic as expected.  This could be attributed to the misalignment of the magnetic moment already detected by SANS measurements.

More information on the magnetism of magnetosome chains has been gathered from the theoretical modelling of the hysteresis loops. Previous works on \emph{M. gryphiswaldense} show that the magnetic anisotropy of magnetosome chains can be described as a superposition of the cubic magnetocrystalline anisotropy of magnetite, with four equivalent easy axes directed along the $\langle 111 \rangle$ crystallographic directions,  and a uniaxial shape anisotropy directed along the chain axis (thus parallel to one of the $\langle 111 \rangle$ axes) that originates from intra-chain dipolar interactions and dominates the magnetic response of the chain~\cite{Charilaou2011,Charilaou2011b}.

Following this argumentation, as a first approximation, our approach to simulate theoretically the experimental hysteresis loops consists of considering the magnetosome chain as a collection of independent magnetic dipoles, where the equilibrium orientation of each magnetic moment is obtained by minimizing the single dipole energy density $E$, calculated as the sum of three contributions: i) cubic magnetocrystalline energy of magnetite with anisotropy constant $K_c$, ii) uniaxial anisotropy energy along the chain axis ([111] direction defined as $\hat{u}_{uni}=\hat{u}_{111}$ in Fig.~\ref{fig:cryotomo2}d) with anisotropy constant $K_{uni}$, and iii) Zeeman energy in an external magnetic field $\mu_0\vec{H}$:
\begin{equation}\label{eq:energy-anis-cub-111}
  E=E_{cubic} + K_{uni} [1-(\hat{u}_{uni}\cdot \hat{u}_m)^2 ] -\mu_0MH(\hat{u}_H \cdot \hat{u}_m)
\end{equation}
where  $\hat{u}_m$ and $\hat{u}_H$ represent the particle magnetization and external magnetic field unit vectors, respectively.  The azimuthal angle can be any between 0\degree\ and 360\degree\, either because individual dipoles are actually free to be rotated along the chain axis  or because whole bacteria themselves can be found at any azimuthal orientation relative to the aligning field due to the sample preparation. The hysteresis loops have then been calculated following a dynamical approach in which the single domain magnetization can switch between the available energy minima states at a rate determined by a Boltzmann factor~\cite{Geoghegan1997,Carrey2011} that depends on the energy barriers between such minima. More details are given in the supplementary information.

Misalignments of the chains with respect to the aligning field occurring during sample preparation and detected by SANS experiments have been considered by including a Gaussian angular distribution of the chain  axes.
We have tested three angular distributions around the chain  axis: $15\degree$, $25\degree$, and $35\degree$, in agreement with misalignment values reported previously \cite{Li2013,Kornig2014}.  The data used for the simulations are the magnetocrystalline anisotropy constant ($K_c=\unit[-11]{kJ/m^3}$) and magnetization ($M_s=\unit[48\cdot 10^4]{A/m}$) of magnetite and an effective uniaxial anisotropy constant along the chain  axis $K_{uni}=\unit[12]{kJ/m^3}$ to account for both shape and magnetic interaction anisotropies. Even though the simulations are good enough at the perpendicular orientation, that is to say, with the applied magnetic field perpendicular to the bacterial chain axis, they clearly deviate at smaller angles as shown in the polar representation in Fig.~\ref{fig:SQUID}c, where the experimental and calculated reduced remanent magnetization and coercivity of bacteria in 2D and 3D arrangements are plotted for different orientation angles between $0$ and $90\degree$.

Since the misalignment of the bacteria due to sample preparation,  does not reproduce the hysteresis loops,  we have tried a different approach in which we have considered that there is a tilting of the magnetization with respect to the chain  axis inherent to the magnetosome. This approach allows reproducing accurately the experimental hysteresis loops for both 2D and 3D arrangements by setting the uniaxial easy axis $\hat{u}_{uni}$ at $25\degree$ with the chain axis in eq.~\ref{eq:energy-anis-cub-111}, in the plane containing the chain axis ([111]) and the [100]  directions. By setting this angle, the effective easy axis, and as a consequence the equilibrium magnetic moment, is found to lie $20\degree$ out of the chain axis, in close agreement with the $25\degree$ tilting observed in the previous SANS analysis of the colloid.
Note that there is a good resemblance with experiment even for an angular dispersion as small as $15\degree$ (Figs.~\ref{fig:SQUID}a,b). We used the same data for the simulations of the hysteresis loops at all orientation angles and for both 3D and 2D arrangements, $K_c=-11$ kJ/m$^3$, $M_s=\unit[48\cdot 10^4]{A/m}$ and $K_{uni}=\unit[12]{kJ/m^3}$. The resemblance with experiment is more evident in the polar plots of the reduced remanent magnetization and coercivity shown in Fig.~\ref{fig:SQUID}c.

The tilting of the magnetosome magnetization with respect to the chain axis is attributed to the competition between the shape anisotropy of the magnetosome, which presents a well faceted morphology as shown in Fig.~\ref{fig:cryotomo2}c, and the magnetic interactions trying to align the magnetic dipoles along the [111] direction parallel to the chain axis. In fact,  electron holography experiments on individual magnetosomes from \emph{Magnetovibrio blakemorei} MV-1 clearly show that the magnetization direction of the particle is tilted with respect to the [111] crystallographic direction towards a long dimension of the particle, consistent with shape anisotropy dominating the magnetic state of the crystal~\cite{Thomas2008}.

\subsubsection*{X-ray photoemission electron microscopy (XPEEM) on extracted magnetosomes}
\label{XPEEM}

Previous magnetic results are supported by x-ray photoemission electron microscopy (XPEEM)~\cite{Kronast2016}. Unlike the macroscopic SQUID and VSM measurements above, which provide an average measurement of the whole sample, XPEEM is an element-specific and spatially-resolved technique that by using x-ray magnetic circular dichroism (XMCD) as a magnetic contrast mechanism, allows obtaining element-specific magnetic hysteresis loops of selected sample areas with a resolution down to 30 nm~\cite{Kronast2016}.

\begin{figure}
\centering
\includegraphics[clip,scale=1]{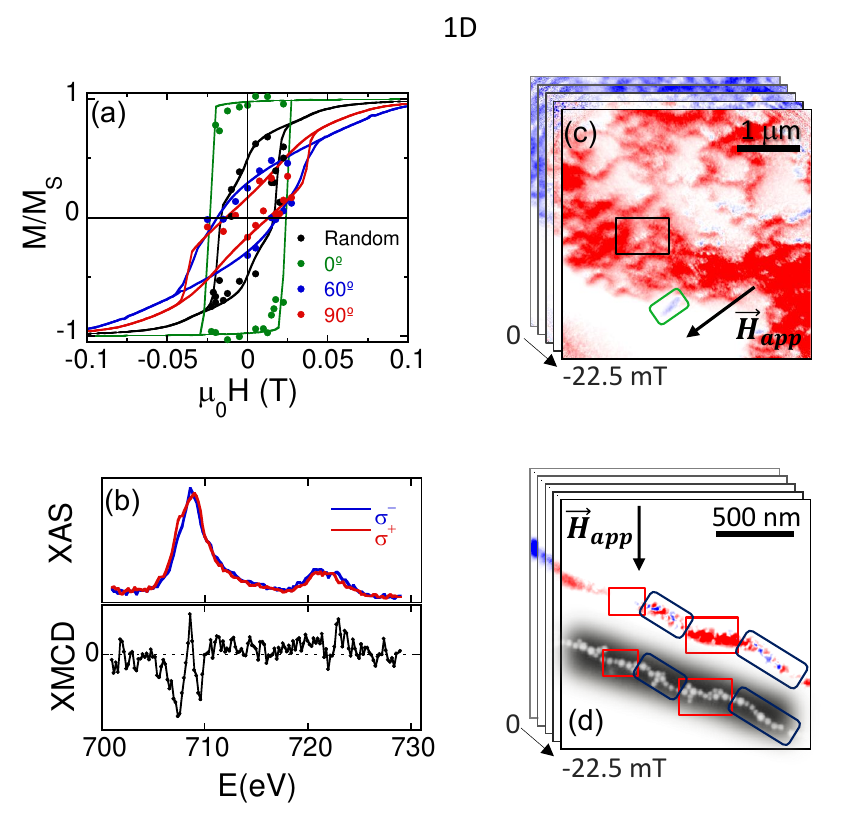}\\
\caption{
a) X-ray photoemission electron microscopy (XPEEM) hysteresis loops of the regions enclosed in the rectangles marked in the images in c) and d), corresponding to chain sections which are either randomly arranged or forming 0\degree\ (green), 60\degree\ (blue), and 90\degree\ (red) with the applied field. The continuous line is the fit to the magnetic model explained in the text.
b) X-ray absorption spectra (XAS) of the chain parallel to the applied field (green rectangle in c) with the incoming beam right-polarized ($\sigma^+$) and left-polarized ($\sigma^-$). Computing $\sigma^+ - \sigma^-$ gives the XMCD signal below.
c) XPEEM images as a function of the applied field of randomly arranged magnetosomes and d) oriented magnetosomes deposited under $\vec{H}_{al}$. In d) the SEM image of the chain is shown together with the XPEEM image.
}\label{fig:XPEEM}
\end{figure}

XPEEM measurements have been performed at room temperature on extracted magnetosomes due to the high absorbing power of the whole bacteria. Magnetosomes were deposited onto a conductive Si substrate either randomly (Fig.~\ref{fig:XPEEM}c) or  oriented along an aligning field (Fig.~\ref{fig:XPEEM}d).
As expected for magnetite, the L$_3$ XMCD signal shows three peaks (Fig.~\ref{fig:XPEEM}b): two minima at $\unit[707.4]{eV}$ and $\unit[709.6]{eV}$ and a maximum at $\unit[708.6]{eV}$. The two minima correspond to the Fe$^{2+}$ and Fe$^{3+}$ ions occupying tetrahedral sites, and the maximum corresponds to the Fe$^{3+}$ ions occupying octahedral sites. The magnetic contrast images were recorded at the Fe $L_3$ resonance energy ($\unit[709.6]{eV}$) of magnetite.
The set of magnetic images displayed in Figs.~\ref{fig:XPEEM}c,d show the space-resolved dichroic images obtained at different values of an in-plane magnetic field. The dichroic signal yields the magnetic moment, so that the hysteresis loops of selected areas of the magnetic images (enclosed in rectangles in Figs.~\ref{fig:XPEEM}c,d) have been obtained by plotting the corresponding dichroic signal as a function of the applied magnetic field (Fig.~\ref{fig:XPEEM}a).

The hysteresis loops of the randomly deposited magnetosomes enclosed in the region marked with the black rectangle in  Fig.~\ref{fig:XPEEM}c and of a chain parallel to the applied field, marked with the green rectangle in Fig.~\ref{fig:XPEEM}c, are shown in Fig.~\ref{fig:XPEEM}a.

Magnetosomes deposited under an applied aligning field form longer chains. The XPEEM image together with the SEM image of one of these chains is shown in Fig.~\ref{fig:XPEEM}d. This chain is clearly not a straight line but is rather a zigzag, formed by segments oriented at different angles with the applied field. Two of these segments, enclosed in blue rectangles in Fig.~\ref{fig:XPEEM}d, form $60\degree$ with the applied field, and another two segments (red rectangles), form $90\degree$. The corresponding hysteresis loops are shown in Fig.~\ref{fig:XPEEM}a.

As shown in Fig.~\ref{fig:XPEEM}a, the experimental XPEEM loops can be accurately simulated to the theoretical model developed previously for the 3D and 2D chain arrangements (eq.~\ref{eq:energy-anis-cub-111}) with a smaller tilting angle of the effective easy axis ($15 \degree$) and a larger anisotropy constant $K_{uni}=\unit[16]{kJ/m^3}$ as compared to the SQUID loops ($K_{uni}=\unit[12]{kJ/m^3}$). This is attributed to  the larger distance between magnetosomes in chains inside the bacteria ($d=\unit[60]{nm}$) than in chains of extracted magnetosomes ($d=\unit[50]{nm}$). Indeed, for a $\unit[40]{nm}$ sized particle, the dipole pair potential energy is given by $\sim \mu_0m^2/4\pi d^3=\unit[0.75]{eV}$ (being $m=M_sV$ the particle magnetic moment and $d=\unit[60]{nm}$), while for the chains of extracted magnetosomes the dipolar energy from the neighbors at $d=\unit[50]{nm}$ would increase up to $\sim \unit[1.35]{eV}$.

\subsection*{Equilibrium configuration of the chain}\label{chain}

\begin{figure}
\centering
\includegraphics[scale=0.7]{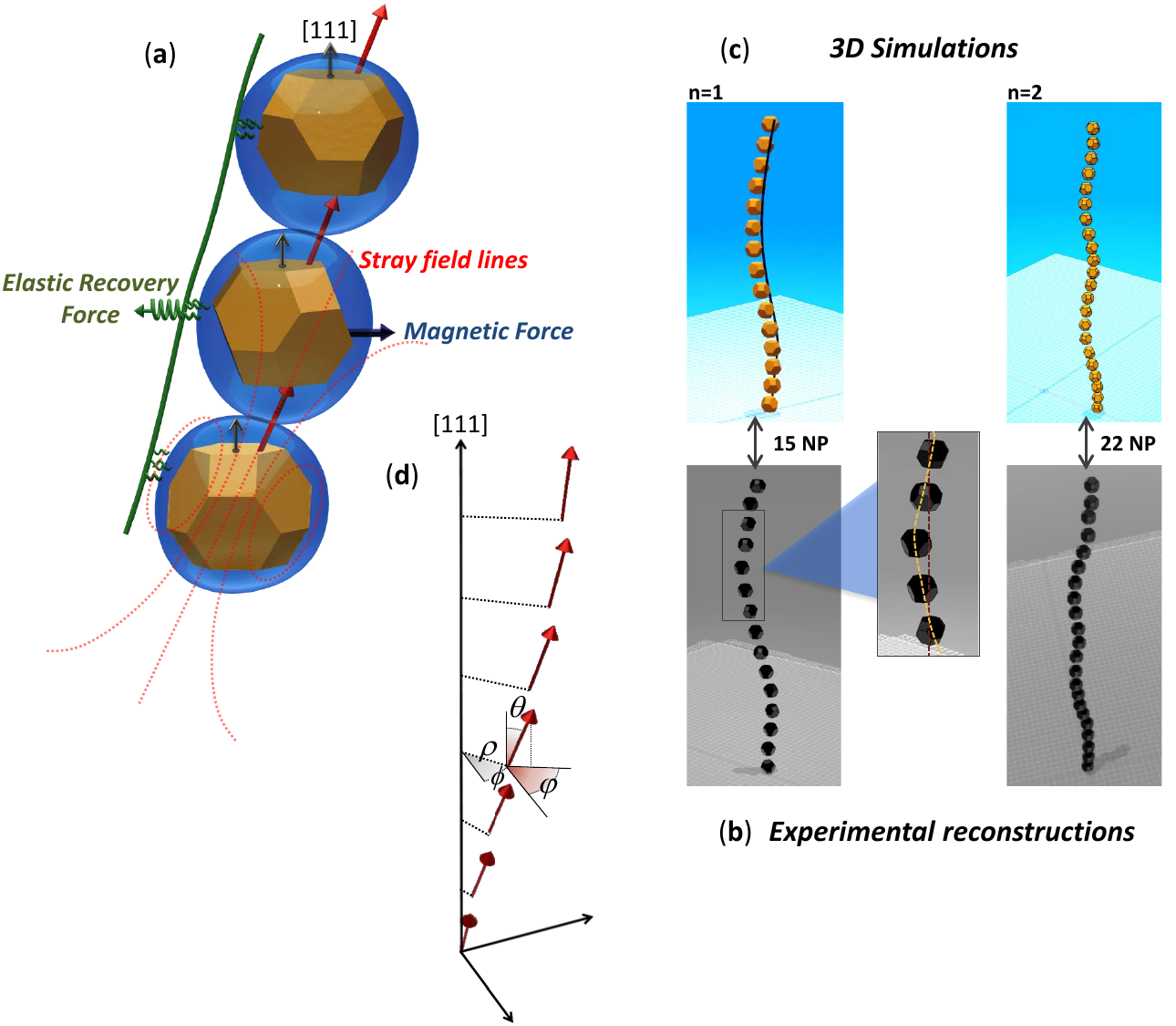}\\
\caption{ Equilibrium configuration of the magnetosome chain.
a) Schematic representation of two competing mechanisms: magnetic force pushing to align magnetosome magnetic moments along the stray field lines from neighboring particles, and lipid/protein-based mechanism modeled as an elastic recovery force acting perpendicularly to the chain axis, where the chain axis is the [111] crystallographic direction along which magnetosomes align, as highlighted in the figure.
b) Experimental reconstructions obtained from ECT imaging of the magnetosome chains shown in Fig.~\ref{fig:cryotomo2}a,b. A zoom-in of the first magnetosome chain reconstruction highlights the deviation from a straight line.
c) Two stable solutions for the chain patterns obtained as explained in the text. A potential filament has been drawn as a guide for the eye. A video of the experimental reconstruction of the chain on the left (chain (a) in Fig.~\ref{fig:cryotomo2}) together with the corresponding simulated chain viewed from different perspectives can be found in the supplementary information (movie S3).
d) Schematic representation of the magnetic dipoles and the three independent variables used in the simulation: radial ($\rho$) and azimuthal ($\phi$)  coordinates for the magnetosome positions, and azimuthal orientation ($\varphi$) of the magnetic dipoles. $\theta$ is the polar angle of the magnetic dipoles, fixed to 20\degree.}
\label{fig:chain_config}
\end{figure}

Following the results gathered from the magnetic analysis, here we will assess the impact on the magnetosome chain configuration of the tilting of the magnetosome magnetization with respect to the chain axis, where the latter, as noted above, is defined as the [111] crystallographic direction along which magnetosomes align in the chain (Fig.~\ref{fig:chain_config}).

With this aim, we have developed an approach to explain the shape of the magnetosome chains that consists on quantifying the total energy of the chain by including the magnetostatic interactions between nanoparticles, and the contribution of the lipid/protein-based architecture embedding the magnetosome chain, modeled as a spring-like elastic energy. We focus on the stable geometry of chain arrangements and assume that close loop configurations such as rings or 3D clusters are avoided by the cytoskeleton inside bacteria. The same approach has been used by other authors~\cite{Kiani2015,Shcherbakov1997}, but  in the present case, as proposed previously, the magnetization of each magnetosome is tilted $20\degree$ out of the chain axis.

Here we will summarize the main steps followed to calculate the total energy of the chain, but a detailed description can be found in the supplementary information. Fig.~\ref{fig:chain_config}a shows a section of a magnetosome chain composed of three magnetosomes. The particle in the centre is subjected to the stray magnetic field produced by the two neighbors (the dotted red lines in Fig.~\ref{fig:chain_config}a represent the stray field produced by the particle at the bottom). We implicitly assume that the local torque that tends to align neighboring dipoles is counter-balanced by the cytoskeleton, since cryotomography and TEM images reflect that the hexagonal faces are face to face~\cite{Mann1984,Kornig2014}. The particle in the middle will tend to align its magnetization along the stray field lines and will undergo a magnetic force that compels it to shift towards the direction of its own magnetic moment. The magnetostatic energy associated to the magnetic force on magnetosomes is then implemented as the sum of the dipolar pair potential energies between nearest neighboring particles.

On the other hand, magnetite crystals are embedded in a lipid/protein-based architecture composed, among others, of the magnetosome membranes, the cytoskeletal filaments, and the proteins connecting the magnetosome membranes with the cytoskeleton. The contribution to the total chain energy of this lipid/protein-based architecture has been modeled based on two simplifying assumptions: firstly, that the distance between particles, $d$, is fixed, so that chains can bend or twist but cannot stretch; and secondly, that the net force on the magnetosomes can be ascribed to an elastic recovery force with elastic constant $k$ acting perpendicularly to the chain  axis, as shown in Fig.~\ref{fig:chain_config}a.

The most stable chain configurations have then been predicted  by minimizing the total energy of the chain considering three independent variables per particle, namely the radial ($\rho$) and azimuthal ($\phi$)  coordinates for the magnetosome positions, and azimuthal orientation ($\varphi$) of the magnetic dipoles (see sketch in Fig.~\ref{fig:chain_config}d). The results have then been compared to the 3D reconstructions obtained from the chains imaged by ECT (Fig.~\ref{fig:chain_config}b).

Considering a particle diameter $D=40$ nm, distance between them $d=60$ nm, and $k=\unit[0.1\cdot 10^{-3}]{N/m}$, two stable solutions give the patterns shown in Fig.~\ref{fig:chain_config}c, in excellent match with the two experimental chain reconstructions shown in Fig.~\ref{fig:chain_config}b. They are helical-like shaped chains with slowly bending axes and a different pitch, given by $\approx 2L/n$, being $L=Nd$ the chain length and $n=1$, and 2, respectively. The energy difference between the two chain shapes is less than 1\% of the total energy, hence they are approximately equally stable. The agreement between the simulated and the experimental chains is even more explicit in a video of images of both chains taken from different perspectives shown in the supplementary information (movie S3).

A zoom-in of a section of one of the simulated chains shows the magnetosome dipole moments. The azimuthal increment of the individual dipole orientations ($\Delta \varphi$) between consecutive positions along the chain increases with $n$ approximately as:
\begin{equation}\label{eq:phi}
  \Delta \varphi \approx 2\pi \frac{n}{N}
\end{equation}
Where $N$ is the number of magnetosomes per chain ($N=15$ and $22$ in the two examples shown). Consequently, the component of the chain magnetization vector (the vector sum of the magnetic moment of each magnetosome) perpendicular to the chain axis cancels out, so that the chain magnetization lies along the chain  axis. Note also that when magnetic dipoles are projected in a 2D view of the chain, apparent tilting between consecutive dipoles is no more than $7\degree$ (see Fig.~4 in the supplementary information), which is very compatible with projected electron holography images~\cite{DuninBorkowski1998}.

The role of $k$ on the chain geometry is to scale up or down the radial positions of magnetosomes along the chain,  so that larger $k$ favours configurations approaching straight lines.  In our case, $k=\unit[0.1\cdot 10^{-3}]{N/m}$ results in assembling forces of the order of 0.6 pN for radial displacements of $\approx \unit[10]{nm}$ as obtained from the simulations, similar to the force generated by the actin filament~\cite{Footer2007}, and  far below recent estimations of the fracture limit of the actin-based scaffolding filaments ($\approx 30$ pN) by K\"{o}rnig \emph{et al}.~\cite{Kornig2014b}.

On the basis of our results, a mechanism of chain formation is proposed. Our findings reveal that the chain shape is mostly driven by passive spontaneous magnetostatic effects triggered by the intrinsic anisotropy of magnetosomes, ultimately defined by their morphology. The magnetosomes morphology is regulated during the biomineralization process. This is a genetically controlled process which involves a specific set of about $30$ \emph{mam} (magnetosome membrane) and \emph{mms} (magnetic particle-membrane specific) genes~\cite{Uebe2016}. The product of the \emph{mamJ} gene is an acidic protein that connects the magnetosome membrane to the filament. The deletion of \emph{mamJ} results in bacteria that produce magnetosomes arranged in compact three-dimensional clusters instead of arranged in chains~\cite{Scheffel2006}. The fact that they form clusters and not closed rings is consistent with our conclusions, namely the tilting of the magnetosome magnetic moment with respect to the chain [111] axis, since according to previous works, more than four magnetosomes tend to form closed rings when their magnetic moment is parallel to the [111] axis~\cite{Kiani2015}. In the same way, when depositing the magnetosomes onto a 2D surface, magnetosomes tend to self-assemble in a close-packing configuration~\cite{Huizar-Felix2016}, and the tilting could be also behind the zigzag configuration of the chains observed by SEM in oriented magnetosomes.
But, why would this helical-like shape benefit the bacteria? \emph{M. gryphiswaldense} are long cells, easily reaching several microns long. They need a high magnetic torque to overcome the drag forces and orientate along the Earth's magnetic field. As a consequence, their magnetosome chain needs to be long to maximize the chain net magnetic moment. Indeed, their chain is frequently composed of more than 20 magnetosomes, which brings the total length of the chain to $\unit[1.5]{\mu m}$ or more, about 50--60\% of the bacterial length. Such an object should necessarily be bent in order to accommodate to this spiral-shaped microorganism. Thus a helical-like shape fits better, but it only changes slightly the total magnetic moment, hence hardly affecting the magnetic orientation. A genetic control of the magnetosome shape towards an energetically optimum chain arrangement is also observed in \emph{Magnetobacterium bavaricum} MYR-1, a species which synthesize bullet-shaped magnetosomes arranged into bundles of magnetosome chains. The magnetosome  magnetization within MYR-1 magnetosomes is parallel to the chain axis, which coincides with the long axis of the magnetosomes. Unexpectedly, the latter is parallel to the  [100] crystallographic axis, a magnetically hard axis in magnetite, rather than
along the [111] magnetocrystalline easy axis, due to compromise effects of shape anisotropy and intra-chain and intra-bundle interactions~\cite{Li2010}.

In sum, our finding sheds light on the  understanding of the magnetosome chain assembly during the biomineralization process of MTB, which may influence their potential future applications as biological micro-robots. Indeed, one of the major technical
issues in the development of interventional platforms for guiding drug-loaded MTB is the directional magnetic field strength that needs to be produced at the human scale to induce sufficient directional torque on the chain of magnetosomes. A good knowledge of the magnetic configuration of the magnetosome chain will lead to more efficient and less costly drug-delivery platforms that may benefice a larger population.

\section*{Methods}

\subsection*{Magnetotactic bacteria culture and magnetosome isolation}

\emph{Magnetospirillum gryphiswaldense} MSR-1 (DMSZ 6631) was grown at 28\celsius\ without sha\-king in an iron rich medium~\cite{Heyen2003}. After $\unit[96]{h}$-incubation, when bacteria present well-formed magnetosomes chains, cells were fixed with $2\%$ glutaraldehyde, harvested by centrifugation, washed three times and finally  concentrated up to $\unit[10^9-10^{11}]{cell/mL}$ in ultrapure water.
Magnetosomes were isolated according to the protocol described by Gr\"{u}nberg et al.~\cite{Grunberg2001} with minor modifications. Cells were collected by centrifugation, suspended in $\unit[20]{mM}$ HEPES-$\unit[4]{mM}$ EDTA (pH=7.4), and disrupted using a French press ($\unit[1.4]{kbar}$). Then, the magnetosomes were collected from the supernatant by magnetic separation and rinsed 10 times with $\unit[10]{mM}$ mM Hepes-$\unit[200]{mM}$ NaCl (pH=7.4). Finally, the isolated magnetosomes were suspended in ultrapure water reaching a concentration of $\unit[20]{\mu g/mL}$.

\subsection*{Electron cryotomography (ECT)}
Sample preparation for cryotomography was as follows: a $\unit[10]{\mu l}$ volume of fixed and washed \emph{M. gryphiswaldense} cells ($\unit[10^9]{cel/mL}$) was mixed with a $\unit[3]{\mu l}$ volume of $\unit[10]{nm}$ Au nano-particles (Aurion\circledR BSA gold tracer $\unit[10]{nm}$) (used as fiducial markers). This mixture is frozen-hydrated following standard methods using a Vitrobot Mark III (FEI Inc., Eindhoven, The Netherlands).
In brief, a grid is placed in the controlled environment of the Vitrobot chamber, which is at $4\celsius$ and at a relative humidity of 95\%. An aliquot ($\unit[4]{\mu l}$) of the sample is applied to a glow-discharged grid. After 1 min incubation most of the drop is removed by blotting with two filter papers, to produce a thin liquid film, and rapidly plunged into liquid ethane ($\unit[91]{K}$), previously cooled by liquid nitrogen.
Vitrified grids are stored under liquid nitrogen until cryo-TEM data collection. For cryo-tomographic tilt series acquisition, vitrified grids were cryo-transferred into a $914$ high tilt tomography cryo-holder (Gatan Inc., Warrendale, PA, USA), which is inserted in a JEM-2000FS/CR field emission gun transmission electron microscope (Jeol, Europe, Croissy-sur-Seine, France) operated at $\unit[200]{kV}$. Grids are kept around $\unit[103]{K}$ in the high vacuum of the microscope column containing the sample embedded in a thin layer of glass-like vitreous ice, in a near native-state.
Different single-axis tilt series were collected under low-dose conditions on a UltraScan 4000, 4k$\times$4k CCD camera (Gatan Inc., Pleasanton, CA, USA), over a tilt range of $\pm 64\degree$ with $1.5\degree$ increments and at underfocus values ranging from 5 to $\unit[8]{\mu m}$, using the semi-automatic data acquisition software SerialEM~\cite{Mastronarde2005}.
Tilt-series were collected at a nominal magnification of $25,000\times$ and a binning factor of 2 ($2048 \times 2048$ pixels micrographs), thus producing a pixel size of $\unit[0.95]{nm}$.
The  in-column omega energy filter helped to record images with improved signal-to-noise-ratio by zero-loss filtering with an energy window of $\unit[60]{eV}$ centered at the zero-loss peak. CCD images in each tilt-series were acquired at the same underfocus value and under the same low-dose conditions.
The maximum total dose used for a tilt-series was $\unit[90]{electrons/ \AA^2}$ consisting of about $\unit[1-2]{electrons/ \AA^2}$ for each digital image.

For the alignment and 3D reconstruction of the tilt-series, we used IMOD software~\cite{Kremer1996}. We employed the Au fiducial markers during the alignment process, and 3D reconstruction was carried out by weight back-projection followed by a reconstruction algorithm named Simultaneous Iterative Reconstruction Technique (SIRT). Resulting tomograms were visualized with ImageJ~\cite{Schindelin2015} as a sequence of cross sectional slices in different plane orientations. Tomograms were processed using a median filter and visualized as 3D electron density maps using UCSF Chimera~\cite{Pettersen2004} software.

\subsection*{Scanning Electron Microscopy (SEM)}
SEM imaging was performed on the two magnetosome samples measured in XPEEM (magnetically oriented and randomly arranged). SEM images were collected at 10 kV with a JEOL JSM-7000F equipped with secondary and retro dispersive electron detectors.

\subsection*{Magnetic measurements}

Magnetic measurements on bacteria forming 2D and 3D arrangements were performed on a superconducting quantum interference device magnetometer (SQUID, Quantum Design MPMS-7) in DC mode and on a vibrating sample magnetometer (VSM).  Isothermal magnetization loops were recorded between $\pm \unit[1]{T}$ at $\unit[300]{K}$.

The 2D bacterial configurations were prepared  by depositing five $\unit[5]{\mu L}$ drops of the bacterial suspension ($\unit[10^9]{cell/mL}$) by the drop coating method~\cite{Gojzewski2012,Huizar-Felix2016} onto a Si substrate. For the oriented bacteria, the deposition was done under an external applied magnetic field of $\unit[0.5]{T}$. To obtain homogeneous samples, infrared radiation was used during the deposition aimed at accelerating the drying and minimising the surface tension. The resulting samples were finally oriented at different angles with respect to the applied magnetic field.

The 3D bacterial configurations were prepared by resuspending $\unit[250]{\mu l}$ of a bacterial colloid ($\unit[10^{11}]{cell/mL}$) in $\unit[750]{\mu l}$ of an agar solution (2\% agar and 98\% water) at $80\celsius$ to maintain the solution in a liquid state. To align the bacteria, a  uniform magnetic field of $\unit[1]{T}$ was applied. After 3 minutes, the field was turned off, and the sample was cooled using liquid
nitrogen until the temperature reached around $0\celsius$. This caused the agar to solidify, trapping
the bacteria, and keeping this solid state at room temperature.

\subsection*{Small angle neutron scattering (SANS) and small angle x-ray scattering (SAXS)}
SANS/SAXS experiments were performed on a highly concentrated bacterial colloid \linebreak[4] \mbox{($\unit[6\cdot 10^{11}]{cell/mL}$)} suspended in ultrapure water. The magnetization curve of the colloid was measured on a vibrating sample magnetometer up to an applied field of $\unit[1]{T}$, leaving two minutes between each  measurement to assure thermal equilibrium was achieved at each point.\\
\textit{Small angle X-ray scattering (SAXS) experiments:}
The SAXS data were collected in a Xenocs Nano-InXider,
utilizing a $\unit[40]{\mu m}$ microfocus Cu anode as X-ray source and a multilayer mo\-no\-chro\-ma\-tor to collect only the Cu-K$\alpha$ radiation.
The detector (a Pilatus 3) is at \unit[938]{mm} from the sample, spans an area of $\unit[83.8\times 33.5]{mm^2}$ and a pixel size of $\unit[172\times 172]{\mu m^2}$. 
The colloidal dispersion was filled into a quartz glass capillary and was measured in absence of an externally applied magnetic field.
Via SAXS exclusively the nuclear scattering intensity $I(\vec{q})\propto |\tilde{N}|^2$ is determined, with $\vec{q}$ being the scattering wave vector and $\tilde{N}(\vec{q})$ being the Fourier transform of the nuclear scattering length density $\rho(\vec{r})$.
To analyze the data we performed an indirect Fourier transform~\cite{glatter1977new} of the radially averaged 1D intensity $I(q)$ to extract the pair distance distribution function $P(r)$.
The real space function $P(r)$ provides direct information about the distances between scatterers from the scattering sample~\cite{glatter1979interpretation} and hence contains information about the average particle geometry as well as correlations between neighboring particles.
For the indirect Fourier transform we used an approach based on ref. ~\cite{Vestergaard:wf5022} and computational details can be found on ref.~\cite{Bender2017}.\\
\textit{Polarized small angle neutron scattering (SANS) experiments:}
SANS instrument D33 at the Institut Laue Langevin (ILL, Grenoble, France)~\cite{Dewhurst2016,Bender2017b} was employed in order to get a longitudinal neutron-spin analysis (POLARIS)~\cite{honecker2010}.
A homogeneous magnetic field $\vec{H}$ was applied perpendicular to the neutron beam ($\vec{H}\perp \vec{k}$) with field amplitudes of $\mu_0H=\unit[2]{mT}$ and \unit[1]{T}.
The mean wavelength of the neutrons was $\lambda=\unit[0.6]{nm}$, with a wavelength spread of $\Delta\lambda/\lambda=10\%$.
The scattering intensities were measured at two different detector distances ($\unit[3]{m}$ as well as $\unit[13.4]{m}$) giving a $q$-range of about $\unit[0.05-0.5]{nm^{-1}}$.

Application of the POLARIS mode enabled us to detect the non spin flip (nsf) intensities $I^{++}(\vec{q})$, $I^{--}(\vec{q})$, where the superscripts indicate the polarization of the incoming neutron beam and the scattered neutrons with regard to the applied field direction, respectively ("+": parallel, "-": antiparallel).
Defining $x$ as the direction of the neutron beam and $z$ as the direction of the applied magnetic field at the sample position ($\vec{H}\perp \vec{k}$) the nsf intensities can be written as~\cite{honecker2010}:
\begin{align}\label{Eq1}
I^{\pm\pm}(\vec{q})\propto&|\tilde{N}|^2+b_H^2|\tilde{M}_z|^2\mathrm{sin}^4\Theta\\
&+b_H^2|\tilde{M}_y|^2\mathrm{sin}^2\Theta\mathrm{cos}^2\Theta\nonumber\\
&-b_H^2(\tilde{M}_y\tilde{M}_z^*+\tilde{M}_z\tilde{M}_y^*)\mathrm{sin}^3\Theta\mathrm{cos}\Theta\nonumber\\
&\mp b_H(\tilde{N}\tilde{M}_z^*+\tilde{N}^*\tilde{M}_z)\mathrm{sin}^2\Theta\nonumber\\
&\pm b_H(\tilde{N}\tilde{M}_y^*+\tilde{N}^*\tilde{M}_y)\mathrm{sin}\Theta\mathrm{cos}\Theta\nonumber.
\end{align}
Here, $\Theta$ is the angle between the scattering vector $\vec{q}$ and the magnetic field $\vec{H}$ and the terms $\tilde{M}_{y,z}(\vec{q})$ represent the Fourier transforms of the magnetization in ${y,z}$ direction.
The superscript $*$ indicates the complex-conjugated quantities and the constant $b_H=2.7\cdot10^{-15}\,\mathrm{m}/\mu_B$, with $\mu_B$ being the Bohr magneton.
To investigate the alignment of the bacteria in field direction we analyzed the 1D nuclear cross sections $I_{nuc}(q)\propto|\tilde{N}|^2$ and the 1D nuclear magnetic cross-terms $I_{cross}(q)\propto(\tilde{N}\tilde{M}_z^*+\tilde{N}^*\tilde{M_z})$.
The purely nuclear scattering intensities were determined by integration of the nsf intensities in $10^{\circ}$ sectors around $\Theta= 0^{\circ}$ ($\vec{q}\parallel\vec{H}$, Eq.\,\ref{Eq1}) and the cross terms $I_{cross}(q)$ by integration of $I^{--}(\vec{q})-I^{++}(\vec{q})$ in $10^{\circ}$ sectors around $\Theta= 90^{\circ}$ ($\vec{q}\perp\vec{H}$, Eq.\,\ref{Eq1}). The difference between the two nsf  cross sections yields information on the polarization-dependent nuclear-magnetic terms. This difference allows one to highlight weak magnetic contributions relative to strong
nuclear scattering (or vice versa).
From the determined cross sections we extracted the underlying pair distance distribution functions in the same manner as for SAXS. In \ce{H2O}, there is contrast match of bacteria in the dispersion, such that only the magnetosomes and their arrangement are probed by SANS.

\subsection*{X-ray photoelectron emission microscopy (XPEEM)}
For the x-ray photoemission electron microscopy (XPEEM) experiments isolated magnetosomes extracted from the bacteria were employed. Two different samples were prepared. The first one consists of randomly arranged magnetosomes prepared by the deposition of a $\unit[5]{\mu L}$ drop of the magnetosomes suspension onto a conductive Si substrate. The drop was dried under infrared radiation. The second sample consists of magnetically oriented magnetosomes. In this case, the drop was dried under an external magnetic field of $\unit[0.5]{T}$ and infrared radiation. The Si substrate was marked with a Au reticule to allow subsequent match by scanning electron microscopy (SEM) of the chains imaged by XPEEM.

Magnetic imaging of the isolated magnetosomes was performed at room temperature by means of photoelectron emission microscopy (PEEM) by using X-ray magnetic circular dichroism (XMCD) as magnetic contrast mechanism. Measurements were carried out at the \linebreak[4] UE49\_PGM SPEEM beamline at Helmholtz-Zentrum Berlin. A magnetic solenoid attached to the sample holder allowed application of an in-plane pulsed magnetic field ranging $\pm \unit[0.1]{T}$ to saturate the sample. During imaging the magnetic field range was restricted to $\pm \unit[27.5]{mT}$. The incoming photon energy was tuned to the Fe L$_3$ resonance ($\unit[709.6]{eV}$) to obtain element specific XMCD images of the magnetosomes as a function of the applied magnetic field. At each field a sequence of images was acquired with incoming circular polarized radiation (90\% of circular photon polarization) with left ($\sigma^-$) and right helicity ($\sigma^+$), respectively. To improve statistics we acquired up to $400$ images per helicity and field ($\unit[3]{s}$ of exposure time). After normalization to a bright-field image, the sequence was drift-corrected, and frames recorded with same helicity were averaged.\\
\emph{XMCD images}: Displayed XMCD images were obtained by computing XMCD = ($(\sigma^+ - \sigma^-)/(\sigma^+ + \sigma^-)$). Due to the low signal, the XMCD strength at the regions of interest was comparable to the noise level at nearby regions with no magnetic particles. In order to enhance the magnetic contrast, for visualization purposes only, the XMCD images have been multiplied by the X-ray absorption image, i.e. XAS$= \sigma^+ + \sigma^-$ after background subtraction.\\
\emph{Space-resolved magnetic hysteresis loops}: The data analysis allowed obtaining the magnetic hysteresis loop of any region within the field of view. In order to obtain the magnetic hysteresis (XMCD vs magnetic field) of a selected magnetosome region we computed the intensity as a function of field and helicity ($\sigma^+$ and $\sigma^-$) on the selected area. A local background subtraction was performed by calculating the intensity variation observed within nearby regions with no magnetic particles (substrate). The XMCD is then calculated as the difference of background corrected $\sigma^+$ and $\sigma^-$ divided by their sum, i.e. XMCD = $(\sigma^+ - \sigma^-)/(\sigma^+ + \sigma^-)$.\\
\emph{Space-resolved XAS and XMCD spectra}: For both, left and right helicities of the incoming circular polarized radiation, five stack of images were obtained as the incoming photon energy crossed the Fe L$_{3,2}$ edges. The photon energy was varied between $\unit[690]{eV}$ and $\unit[730]{eV}$ in $\unit[0.2]{eV}$ steps. Total integration time per energy was $\unit[2.5]{s}$. After normalization to a bright-field image, the sequence was drift-corrected, and frames recorded with same helicity and photon energy were averaged. Computing of local X-ray absorption spectra (XAS) for $\sigma^+$ and $\sigma^-$ allows calculation of the X-ray magnetic circular dichroism spectrum as $\sigma^+ - \sigma^-$ from a selected region.

\section*{Acknowledgments}

The Spanish Government is
acknowledged for funding under project number MAT2014-55049-C2-R. The Basque Government is acknowledged for L.M.'s fellowship and for funding under project number IT711-13. We also acknowledge funding from the EU through project NanoMag (grant agreement no. 604448). We thank D. Mu\~noz, D. Gandia and A. Vigilante for preparing the samples for the magnetometry measurements.

\section*{Author contributions}
 I.O. and M.F.G. designed research;
 I.O. and L.M. performed and analyzed SQUID and VSM measurements;
 L.M., A.G.P., A.M., M.F.G, S.V. and M.A.A. performed XPEEM measurements;
 L.M., S.V. and A.G.P. analyzed XPEEM data;
 P.B, L.M., D.A.V. and D.H. performed SAXS/SANS experiments;
 P.B. analyzed SAXS/SANS data;
 I.O. conceived the chain configuration model;
 I.O., L.M., P.B., A.G.P., A.G.A., L.F.B and M.F.G. discussed the magnetic results;
 L.M., A.M., M.F.G. and D.G. performed and analyzed the cryotomography imaging;
 I.O., L.M., P.B., A.G.P. and M.F.G. wrote the paper;
 all authors checked the manuscript;
 A.M. and M.F.G. supervised the work.

\section*{Conflict of interest}

There are no conflicts to declare.

\section*{Electronic Supplementary Information}

Electronic Supplementary Information (ESI) available:
\begin{itemize}
  \item TEM image of oriented bacteria, details on the calculation of the hysteresis loops, calculation of the chain energy, 2D projection of the magnetic dipoles.
  \item Movies S1-S2: videos of the tomograms of the three chains in Fig.~\ref{fig:cryotomo2}a, b viewed as consecutive slices in XY, YZ, and YZ plane orientation, and as an electron density map.
\item Movie S3: video of the experimental reconstruction of the chain in Fig~\ref{fig:cryotomo2}a (corresponding to $n=1$ in Fig.~\ref{fig:chain_config}b)  together with the corresponding simulated chain (Fig.~\ref{fig:chain_config}c) viewed from different perspectives.
\end{itemize}


\begin{mcitethebibliography}{56}
\providecommand*{\natexlab}[1]{#1}
\providecommand*{\mciteSetBstSublistMode}[1]{}
\providecommand*{\mciteSetBstMaxWidthForm}[2]{}
\providecommand*{\mciteBstWouldAddEndPuncttrue}
  {\def\EndOfBibitem{\unskip.}}
\providecommand*{\mciteBstWouldAddEndPunctfalse}
  {\let\EndOfBibitem\relax}
\providecommand*{\mciteSetBstMidEndSepPunct}[3]{}
\providecommand*{\mciteSetBstSublistLabelBeginEnd}[3]{}
\providecommand*{\EndOfBibitem}{}
\mciteSetBstSublistMode{f}
\mciteSetBstMaxWidthForm{subitem}
{(\emph{\alph{mcitesubitemcount}})}
\mciteSetBstSublistLabelBeginEnd{\mcitemaxwidthsubitemform\space}
{\relax}{\relax}

\bibitem[Blakemore(1975)]{Blakemore1975}
R.~Blakemore, \emph{Science}, 1975, \textbf{190}, 377\relax
\mciteBstWouldAddEndPuncttrue
\mciteSetBstMidEndSepPunct{\mcitedefaultmidpunct}
{\mcitedefaultendpunct}{\mcitedefaultseppunct}\relax
\EndOfBibitem
\bibitem[Dunin-Borkowski \emph{et~al.}(1998)Dunin-Borkowski, McCartney,
  Frankel, Bazylinski, Posfai, and Buseck]{DuninBorkowski1998}
R.~E. Dunin-Borkowski, M.~R. McCartney, R.~B. Frankel, D.~A. Bazylinski,
  M.~Posfai and P.~R. Buseck, \emph{Science}, 1998, \textbf{282}, 1868\relax
\mciteBstWouldAddEndPuncttrue
\mciteSetBstMidEndSepPunct{\mcitedefaultmidpunct}
{\mcitedefaultendpunct}{\mcitedefaultseppunct}\relax
\EndOfBibitem
\bibitem[Bazylinski and Frankel(2004)]{Bazylinski2004}
D.~A. Bazylinski and R.~B. Frankel, \emph{Nat. Rev. Micro.}, 2004, \textbf{2},
  217--30\relax
\mciteBstWouldAddEndPuncttrue
\mciteSetBstMidEndSepPunct{\mcitedefaultmidpunct}
{\mcitedefaultendpunct}{\mcitedefaultseppunct}\relax
\EndOfBibitem
\bibitem[Uebe and Sch{\"{u}}ler(2016)]{Uebe2016}
R.~Uebe and D.~Sch{\"{u}}ler, \emph{Nat. Rev. Microbiol.}, 2016, \textbf{14},
  621--637\relax
\mciteBstWouldAddEndPuncttrue
\mciteSetBstMidEndSepPunct{\mcitedefaultmidpunct}
{\mcitedefaultendpunct}{\mcitedefaultseppunct}\relax
\EndOfBibitem
\bibitem[Lefevre and Bazylinski(2013)]{Lefevre2013}
C.~T. Lefevre and D.~A. Bazylinski, \emph{Microbiol. Mol. Biol. Rev.}, 2013,
  \textbf{77}, 497--526\relax
\mciteBstWouldAddEndPuncttrue
\mciteSetBstMidEndSepPunct{\mcitedefaultmidpunct}
{\mcitedefaultendpunct}{\mcitedefaultseppunct}\relax
\EndOfBibitem
\bibitem[Serantes \emph{et~al.}(2014)Serantes, Simeonidis, Angelakeris,
  Chubykalo-Fesenko, Marciello, Morales, Baldomir, and
  Mart{\'{i}}nez-Boubeta]{Serantes2014}
D.~Serantes, K.~Simeonidis, M.~Angelakeris, O.~Chubykalo-Fesenko, M.~Marciello,
  M.~Morales, D.~Baldomir and C.~Mart{\'{i}}nez-Boubeta, \emph{J. Phys. Chem.
  C}, 2014, \textbf{118}, 5927--5934\relax
\mciteBstWouldAddEndPuncttrue
\mciteSetBstMidEndSepPunct{\mcitedefaultmidpunct}
{\mcitedefaultendpunct}{\mcitedefaultseppunct}\relax
\EndOfBibitem
\bibitem[Felfoul \emph{et~al.}(2016)Felfoul, Mohammadi, Taherkhani, de~Lanauze,
  {Zhong Xu}, Loghin, Essa, Jancik, Houle, Lafleur, Gaboury, Tabrizian, Kaou,
  Atkin, Vuong, Batist, Beauchemin, Radzioch, and Martel]{Felfoul2016}
O.~Felfoul, M.~Mohammadi, S.~Taherkhani, D.~de~Lanauze, Y.~{Zhong Xu},
  D.~Loghin, S.~Essa, S.~Jancik, D.~Houle, M.~Lafleur, L.~Gaboury,
  M.~Tabrizian, N.~Kaou, M.~Atkin, T.~Vuong, G.~Batist, N.~Beauchemin,
  D.~Radzioch and S.~Martel, \emph{Nat. Nanotechnol.}, 2016, \textbf{11},
  941--949\relax
\mciteBstWouldAddEndPuncttrue
\mciteSetBstMidEndSepPunct{\mcitedefaultmidpunct}
{\mcitedefaultendpunct}{\mcitedefaultseppunct}\relax
\EndOfBibitem
\bibitem[Ghosh \emph{et~al.}(2012)Ghosh, Lee, Thomas, Kohli, Yun, Belcher, and
  Kelly]{Ghosh2012}
D.~Ghosh, Y.~Lee, S.~Thomas, A.~G. Kohli, D.~S. Yun, A.~M. Belcher and K.~A.
  Kelly, \emph{Nat. Nanotechnol.}, 2012, \textbf{7}, 677--682\relax
\mciteBstWouldAddEndPuncttrue
\mciteSetBstMidEndSepPunct{\mcitedefaultmidpunct}
{\mcitedefaultendpunct}{\mcitedefaultseppunct}\relax
\EndOfBibitem
\bibitem[Alphand{\'{e}}ry \emph{et~al.}(2011)Alphand{\'{e}}ry, Faure, Seksek,
  Guyot, and Chebbi]{Alphandery2011}
E.~Alphand{\'{e}}ry, S.~Faure, O.~Seksek, F.~Guyot and I.~Chebbi, \emph{ACS
  Nano}, 2011, \textbf{5}, 6279--6296\relax
\mciteBstWouldAddEndPuncttrue
\mciteSetBstMidEndSepPunct{\mcitedefaultmidpunct}
{\mcitedefaultendpunct}{\mcitedefaultseppunct}\relax
\EndOfBibitem
\bibitem[Mishra \emph{et~al.}(2016)Mishra, Dickey, Velev, and
  Tracy]{Mishra2016}
S.~R. Mishra, M.~D. Dickey, O.~D. Velev and J.~B. Tracy, \emph{Nanoscale},
  2016, \textbf{8}, 1309--1313\relax
\mciteBstWouldAddEndPuncttrue
\mciteSetBstMidEndSepPunct{\mcitedefaultmidpunct}
{\mcitedefaultendpunct}{\mcitedefaultseppunct}\relax
\EndOfBibitem
\bibitem[Jiang \emph{et~al.}(2016)Jiang, Feng, Huang, Wu, Su, Yang, Mai, and
  Jiang]{Jiang2016}
X.~Jiang, J.~Feng, L.~Huang, Y.~Wu, B.~Su, W.~Yang, L.~Mai and L.~Jiang,
  \emph{Adv. Mater.}, 2016, \textbf{28}, 6952--6958\relax
\mciteBstWouldAddEndPuncttrue
\mciteSetBstMidEndSepPunct{\mcitedefaultmidpunct}
{\mcitedefaultendpunct}{\mcitedefaultseppunct}\relax
\EndOfBibitem
\bibitem[Kou \emph{et~al.}(2011)Kou, Fan, Dumas, Lu, Zhang, Zhu, Zhang, Liu,
  and Xiao]{Kou2011}
X.~Kou, X.~Fan, R.~K. Dumas, Q.~Lu, Y.~Zhang, H.~Zhu, X.~Zhang, K.~Liu and
  J.~Q. Xiao, \emph{Adv. Mater.}, 2011, \textbf{23}, 1393--1397\relax
\mciteBstWouldAddEndPuncttrue
\mciteSetBstMidEndSepPunct{\mcitedefaultmidpunct}
{\mcitedefaultendpunct}{\mcitedefaultseppunct}\relax
\EndOfBibitem
\bibitem[Komeili(2007)]{Komeili2007}
A.~Komeili, \emph{Annu. Rev. Biochem.}, 2007, \textbf{76}, 351--66\relax
\mciteBstWouldAddEndPuncttrue
\mciteSetBstMidEndSepPunct{\mcitedefaultmidpunct}
{\mcitedefaultendpunct}{\mcitedefaultseppunct}\relax
\EndOfBibitem
\bibitem[Faivre and Sch\"{u}ler(2008)]{Faivre2008}
D.~Faivre and D.~Sch\"{u}ler, \emph{Chem. Rev.}, 2008, \textbf{108},
  4875--4898\relax
\mciteBstWouldAddEndPuncttrue
\mciteSetBstMidEndSepPunct{\mcitedefaultmidpunct}
{\mcitedefaultendpunct}{\mcitedefaultseppunct}\relax
\EndOfBibitem
\bibitem[Komeili \emph{et~al.}(2006)Komeili, Li, Newman, and
  Jensen]{Komeili2006}
A.~Komeili, Z.~Li, D.~K. Newman and G.~J. Jensen, \emph{Science}, 2006,
  \textbf{311}, 242--245\relax
\mciteBstWouldAddEndPuncttrue
\mciteSetBstMidEndSepPunct{\mcitedefaultmidpunct}
{\mcitedefaultendpunct}{\mcitedefaultseppunct}\relax
\EndOfBibitem
\bibitem[Frankel \emph{et~al.}(1983)Frankel, Papaefthymiou, Blakemore, and
  O'Brien]{Frankel1983}
R.~B. Frankel, G.~C. Papaefthymiou, R.~P. Blakemore and W.~O'Brien,
  \emph{Biochim. Biophys. Acta}, 1983, \textbf{763}, 147--159\relax
\mciteBstWouldAddEndPuncttrue
\mciteSetBstMidEndSepPunct{\mcitedefaultmidpunct}
{\mcitedefaultendpunct}{\mcitedefaultseppunct}\relax
\EndOfBibitem
\bibitem[Fdez-Gubieda \emph{et~al.}(2013)Fdez-Gubieda, Muela, Alonso,
  Garc{\'\i}a-Prieto, Olivi, Fern{\'a}ndez-Pacheco, and
  Barandiar{\'a}n]{FdezGubieda2013}
M.~L. Fdez-Gubieda, A.~Muela, J.~Alonso, A.~Garc{\'\i}a-Prieto, L.~Olivi,
  R.~Fern{\'a}ndez-Pacheco and J.~M. Barandiar{\'a}n, \emph{ACS Nano}, 2013,
  \textbf{7}, 3297--305\relax
\mciteBstWouldAddEndPuncttrue
\mciteSetBstMidEndSepPunct{\mcitedefaultmidpunct}
{\mcitedefaultendpunct}{\mcitedefaultseppunct}\relax
\EndOfBibitem
\bibitem[Baumgartner \emph{et~al.}(2013)Baumgartner, Morin, Menguy, {Perez
  Gonzalez}, Widdrat, Cosmidis, and Faivre]{Baumgartner2013}
J.~Baumgartner, G.~Morin, N.~Menguy, T.~{Perez Gonzalez}, M.~Widdrat,
  J.~Cosmidis and D.~Faivre, \emph{PNAS}, 2013, \textbf{1}, 1--6\relax
\mciteBstWouldAddEndPuncttrue
\mciteSetBstMidEndSepPunct{\mcitedefaultmidpunct}
{\mcitedefaultendpunct}{\mcitedefaultseppunct}\relax
\EndOfBibitem
\bibitem[Komeili(2012)]{Komeili2012}
A.~Komeili, \emph{FEMS Microbiol. Rev.}, 2012, \textbf{36}, 232--255\relax
\mciteBstWouldAddEndPuncttrue
\mciteSetBstMidEndSepPunct{\mcitedefaultmidpunct}
{\mcitedefaultendpunct}{\mcitedefaultseppunct}\relax
\EndOfBibitem
\bibitem[Toro-Nahuelpan \emph{et~al.}(2016)Toro-Nahuelpan, M{\"u}ller, Klumpp,
  Plitzko, Bramkamp, and Sch{\"u}ler]{Toro2016}
M.~Toro-Nahuelpan, F.~D. M{\"u}ller, S.~Klumpp, J.~M. Plitzko, M.~Bramkamp and
  D.~Sch{\"u}ler, \emph{BMC Biology}, 2016,  1--24\relax
\mciteBstWouldAddEndPuncttrue
\mciteSetBstMidEndSepPunct{\mcitedefaultmidpunct}
{\mcitedefaultendpunct}{\mcitedefaultseppunct}\relax
\EndOfBibitem
\bibitem[Cornejo \emph{et~al.}(2016)Cornejo, Subramanian, Li, Jensen, and
  Komeili]{Cornejo2016}
E.~Cornejo, P.~Subramanian, Z.~Li, G.~J. Jensen and A.~Komeili, \emph{mBio},
  2016, \textbf{7}, e01898--15\relax
\mciteBstWouldAddEndPuncttrue
\mciteSetBstMidEndSepPunct{\mcitedefaultmidpunct}
{\mcitedefaultendpunct}{\mcitedefaultseppunct}\relax
\EndOfBibitem
\bibitem[Shcherbakov \emph{et~al.}(1997)Shcherbakov, Winklhofer, Hanzlik, and
  Petersen]{Shcherbakov1997}
V.~P. Shcherbakov, M.~Winklhofer, M.~Hanzlik and N.~Petersen, \emph{Eur.
  Biophys. J.}, 1997, \textbf{26}, 319--326\relax
\mciteBstWouldAddEndPuncttrue
\mciteSetBstMidEndSepPunct{\mcitedefaultmidpunct}
{\mcitedefaultendpunct}{\mcitedefaultseppunct}\relax
\EndOfBibitem
\bibitem[Klumpp and Faivre(2012)]{Klumpp2012}
S.~Klumpp and D.~Faivre, \emph{PLoS One}, 2012, \textbf{7}, 1--11\relax
\mciteBstWouldAddEndPuncttrue
\mciteSetBstMidEndSepPunct{\mcitedefaultmidpunct}
{\mcitedefaultendpunct}{\mcitedefaultseppunct}\relax
\EndOfBibitem
\bibitem[Meyra \emph{et~al.}(2016)Meyra, Zarragoicoechea, and Kuz]{Meyra2016}
A.~G. Meyra, G.~J. Zarragoicoechea and V.~A. Kuz, \emph{Phys. Chem. Chem.
  Phys.}, 2016, \textbf{18}, 12768--12773\relax
\mciteBstWouldAddEndPuncttrue
\mciteSetBstMidEndSepPunct{\mcitedefaultmidpunct}
{\mcitedefaultendpunct}{\mcitedefaultseppunct}\relax
\EndOfBibitem
\bibitem[Katzmann \emph{et~al.}(2010)Katzmann, Scheffel, Gruska, Plitzko, and
  Sch\"{u}ler]{Katzmann2010}
E.~Katzmann, A.~Scheffel, M.~Gruska, J.~M. Plitzko and D.~Sch\"{u}ler,
  \emph{Mol. Microbiol.}, 2010, \textbf{77}, 208--224\relax
\mciteBstWouldAddEndPuncttrue
\mciteSetBstMidEndSepPunct{\mcitedefaultmidpunct}
{\mcitedefaultendpunct}{\mcitedefaultseppunct}\relax
\EndOfBibitem
\bibitem[Scheffel \emph{et~al.}(2006)Scheffel, Gruska, Faivre, Linaroudis,
  Plitzko, and Schuler]{Scheffel2006}
A.~Scheffel, M.~Gruska, D.~Faivre, A.~Linaroudis, J.~M. Plitzko and D.~Schuler,
  \emph{Nature}, 2006, \textbf{440}, 110--114\relax
\mciteBstWouldAddEndPuncttrue
\mciteSetBstMidEndSepPunct{\mcitedefaultmidpunct}
{\mcitedefaultendpunct}{\mcitedefaultseppunct}\relax
\EndOfBibitem
\bibitem[Mann \emph{et~al.}(1984)Mann, Frankel, and Blakemore]{Mann1984}
S.~Mann, R.~B. Frankel and R.~P. Blakemore, \emph{Nature}, 1984, \textbf{310},
  405--407\relax
\mciteBstWouldAddEndPuncttrue
\mciteSetBstMidEndSepPunct{\mcitedefaultmidpunct}
{\mcitedefaultendpunct}{\mcitedefaultseppunct}\relax
\EndOfBibitem
\bibitem[K{\"{o}}rnig \emph{et~al.}(2014)K{\"{o}}rnig, Winklhofer, Baumgartner,
  Gonz\'{a}lez, Fratzl, and Faivre]{Kornig2014}
A.~K{\"{o}}rnig, M.~Winklhofer, J.~Baumgartner, T.~P. Gonz\'{a}lez, P.~Fratzl
  and D.~Faivre, \emph{Adv. Funct. Mater.}, 2014, \textbf{24}, 3926--3932\relax
\mciteBstWouldAddEndPuncttrue
\mciteSetBstMidEndSepPunct{\mcitedefaultmidpunct}
{\mcitedefaultendpunct}{\mcitedefaultseppunct}\relax
\EndOfBibitem
\bibitem[Hoell \emph{et~al.}(2004)Hoell, Wiedenmann, Heyen, and
  Sch{\"{u}}ler]{Hoell2004}
A.~Hoell, A.~Wiedenmann, U.~Heyen and D.~Sch{\"{u}}ler, \emph{Physica B}, 2004,
  \textbf{350}, 309--313\relax
\mciteBstWouldAddEndPuncttrue
\mciteSetBstMidEndSepPunct{\mcitedefaultmidpunct}
{\mcitedefaultendpunct}{\mcitedefaultseppunct}\relax
\EndOfBibitem
\bibitem[Li \emph{et~al.}(2013)Li, Ge, Pan, Williams, Liu, and Qin]{Li2013}
J.~Li, K.~Ge, Y.~Pan, W.~Williams, Q.~Liu and H.~Qin, \emph{Geochem. Geophys.
  Geosyst.}, 2013, \textbf{14}, 3887--3907\relax
\mciteBstWouldAddEndPuncttrue
\mciteSetBstMidEndSepPunct{\mcitedefaultmidpunct}
{\mcitedefaultendpunct}{\mcitedefaultseppunct}\relax
\EndOfBibitem
\bibitem[Stoner and Wohlfarth(1948)]{Stoner1948}
E.~C. Stoner and E.~P. Wohlfarth, \emph{Philos. Trans. R. Soc. London, Ser. A},
  1948, \textbf{240}, 599--642\relax
\mciteBstWouldAddEndPuncttrue
\mciteSetBstMidEndSepPunct{\mcitedefaultmidpunct}
{\mcitedefaultendpunct}{\mcitedefaultseppunct}\relax
\EndOfBibitem
\bibitem[Charilaou \emph{et~al.}(2011)Charilaou, Winklhofer, and
  Gehring]{Charilaou2011}
M.~Charilaou, M.~Winklhofer and A.~U. Gehring, \emph{J. Appl. Phys.}, 2011,
  \textbf{109}, 1--6\relax
\mciteBstWouldAddEndPuncttrue
\mciteSetBstMidEndSepPunct{\mcitedefaultmidpunct}
{\mcitedefaultendpunct}{\mcitedefaultseppunct}\relax
\EndOfBibitem
\bibitem[Charilaou \emph{et~al.}(2011)Charilaou, Sahu, Faivre, Fischer,
  Garc\'{\i}a-Rubio, and Gehring]{Charilaou2011b}
M.~Charilaou, K.~K. Sahu, D.~Faivre, A.~Fischer, I.~Garc\'{\i}a-Rubio and A.~U.
  Gehring, \emph{Appl. Phys. Lett.}, 2011, \textbf{99}, 9--11\relax
\mciteBstWouldAddEndPuncttrue
\mciteSetBstMidEndSepPunct{\mcitedefaultmidpunct}
{\mcitedefaultendpunct}{\mcitedefaultseppunct}\relax
\EndOfBibitem
\bibitem[Geoghegan \emph{et~al.}(1997)Geoghegan, Coffey, and
  Mulligan]{Geoghegan1997}
L.~J. Geoghegan, W.~T. Coffey and B.~Mulligan, \emph{Adv. Chem. Phys.}, 1997,
  \textbf{100}, 475 -- 641\relax
\mciteBstWouldAddEndPuncttrue
\mciteSetBstMidEndSepPunct{\mcitedefaultmidpunct}
{\mcitedefaultendpunct}{\mcitedefaultseppunct}\relax
\EndOfBibitem
\bibitem[Carrey \emph{et~al.}(2011)Carrey, Mehdaoui, and Respaud]{Carrey2011}
J.~Carrey, B.~Mehdaoui and M.~Respaud, \emph{J. Appl. Phys.}, 2011,
  \textbf{109}, 083921\relax
\mciteBstWouldAddEndPuncttrue
\mciteSetBstMidEndSepPunct{\mcitedefaultmidpunct}
{\mcitedefaultendpunct}{\mcitedefaultseppunct}\relax
\EndOfBibitem
\bibitem[Thomas \emph{et~al.}(2008)Thomas, Simpson, Kasama, and
  Dunin-Borkowski]{Thomas2008}
J.~M. Thomas, E.~T. Simpson, T.~Kasama and R.~E. Dunin-Borkowski,
  \emph{Accounts of chemical research}, 2008, \textbf{41}, 665--74\relax
\mciteBstWouldAddEndPuncttrue
\mciteSetBstMidEndSepPunct{\mcitedefaultmidpunct}
{\mcitedefaultendpunct}{\mcitedefaultseppunct}\relax
\EndOfBibitem
\bibitem[Kronast and Valencia(2016)]{Kronast2016}
F.~Kronast and S.~Valencia, \emph{JLSRF}, 2016, \textbf{A90}, 1--6\relax
\mciteBstWouldAddEndPuncttrue
\mciteSetBstMidEndSepPunct{\mcitedefaultmidpunct}
{\mcitedefaultendpunct}{\mcitedefaultseppunct}\relax
\EndOfBibitem
\bibitem[Kiani \emph{et~al.}(2015)Kiani, Faivre, and Klumpp]{Kiani2015}
B.~Kiani, D.~Faivre and S.~Klumpp, \emph{New J. Phys.}, 2015, \textbf{17},
  43007\relax
\mciteBstWouldAddEndPuncttrue
\mciteSetBstMidEndSepPunct{\mcitedefaultmidpunct}
{\mcitedefaultendpunct}{\mcitedefaultseppunct}\relax
\EndOfBibitem
\bibitem[Footer \emph{et~al.}(2007)Footer, Kerssemakers, Theriot, and
  Dogterom]{Footer2007}
M.~J. Footer, J.~W.~J. Kerssemakers, J.~A. Theriot and M.~Dogterom,
  \emph{PNAS}, 2007, \textbf{104}, 2181--6\relax
\mciteBstWouldAddEndPuncttrue
\mciteSetBstMidEndSepPunct{\mcitedefaultmidpunct}
{\mcitedefaultendpunct}{\mcitedefaultseppunct}\relax
\EndOfBibitem
\bibitem[K{\"{o}}rnig \emph{et~al.}(2014)K{\"{o}}rnig, Dong, Bennet, Widdrat,
  Andert, M{\"{u}}ller, Sch{\"{u}}ler, Klumpp, and Faivre]{Kornig2014b}
A.~K{\"{o}}rnig, J.~Dong, M.~Bennet, M.~Widdrat, J.~Andert, F.~D. M{\"{u}}ller,
  D.~Sch{\"{u}}ler, S.~Klumpp and D.~Faivre, \emph{Nano Lett.}, 2014,
  \textbf{14}, 4653--4659\relax
\mciteBstWouldAddEndPuncttrue
\mciteSetBstMidEndSepPunct{\mcitedefaultmidpunct}
{\mcitedefaultendpunct}{\mcitedefaultseppunct}\relax
\EndOfBibitem
\bibitem[Hu{\'{i}}zar-F{\'{e}}lix \emph{et~al.}(2016)Hu{\'{i}}zar-F{\'{e}}lix,
  Mu{\~{n}}oz, Orue, Mag{\'{e}}n, Ibarra, Barandiar{\'{a}}n, Muela, and
  Fdez-Gubieda]{Huizar-Felix2016}
A.~M. Hu{\'{i}}zar-F{\'{e}}lix, D.~Mu{\~{n}}oz, I.~Orue, C.~Mag{\'{e}}n,
  A.~Ibarra, J.~M. Barandiar{\'{a}}n, A.~Muela and M.~L. Fdez-Gubieda,
  \emph{Appl. Phys. Lett.}, 2016, \textbf{108}, 063109\relax
\mciteBstWouldAddEndPuncttrue
\mciteSetBstMidEndSepPunct{\mcitedefaultmidpunct}
{\mcitedefaultendpunct}{\mcitedefaultseppunct}\relax
\EndOfBibitem
\bibitem[Li \emph{et~al.}(2010)Li, Pan, Liu, Yu-Zhang, Menguy, Che, Qin, Lin,
  Wu, Petersen, and Yang]{Li2010}
J.~Li, Y.~Pan, Q.~Liu, K.~Yu-Zhang, N.~Menguy, R.~Che, H.~Qin, W.~Lin, W.~Wu,
  N.~Petersen and X.~Yang, \emph{Earth Planet. Sci. Lett.}, 2010, \textbf{293},
  368--376\relax
\mciteBstWouldAddEndPuncttrue
\mciteSetBstMidEndSepPunct{\mcitedefaultmidpunct}
{\mcitedefaultendpunct}{\mcitedefaultseppunct}\relax
\EndOfBibitem
\bibitem[Heyen and Sch\"{u}ler(2003)]{Heyen2003}
U.~Heyen and D.~Sch\"{u}ler, \emph{Appl. Microbiol. Biotechnol.}, 2003,
  \textbf{61}, 536--44\relax
\mciteBstWouldAddEndPuncttrue
\mciteSetBstMidEndSepPunct{\mcitedefaultmidpunct}
{\mcitedefaultendpunct}{\mcitedefaultseppunct}\relax
\EndOfBibitem
\bibitem[Gr{\"{u}}nberg \emph{et~al.}(2001)Gr{\"{u}}nberg, Wawer, and
  Tebo]{Grunberg2001}
K.~Gr{\"{u}}nberg, C.~Wawer and B.~M. Tebo, \emph{Appl. Environ. Microbiol.},
  2001, \textbf{67}, 4573--4582\relax
\mciteBstWouldAddEndPuncttrue
\mciteSetBstMidEndSepPunct{\mcitedefaultmidpunct}
{\mcitedefaultendpunct}{\mcitedefaultseppunct}\relax
\EndOfBibitem
\bibitem[Mastronarde(2005)]{Mastronarde2005}
D.~N. Mastronarde, \emph{J. Struct. Biol.}, 2005, \textbf{152}, 36 -- 51\relax
\mciteBstWouldAddEndPuncttrue
\mciteSetBstMidEndSepPunct{\mcitedefaultmidpunct}
{\mcitedefaultendpunct}{\mcitedefaultseppunct}\relax
\EndOfBibitem
\bibitem[Kremer \emph{et~al.}(1996)Kremer, Mastronarde, and
  McIntosh]{Kremer1996}
J.~R. Kremer, D.~N. Mastronarde and J.~McIntosh, \emph{J. Struct. Biol.}, 1996,
  \textbf{116}, 71 -- 76\relax
\mciteBstWouldAddEndPuncttrue
\mciteSetBstMidEndSepPunct{\mcitedefaultmidpunct}
{\mcitedefaultendpunct}{\mcitedefaultseppunct}\relax
\EndOfBibitem
\bibitem[Schindelin \emph{et~al.}(2015)Schindelin, Rueden, Hiner, and
  Eliceiri]{Schindelin2015}
J.~Schindelin, C.~T. Rueden, M.~C. Hiner and K.~W. Eliceiri, \emph{Mol. Reprod.
  Dev.}, 2015, \textbf{82}, 518--529\relax
\mciteBstWouldAddEndPuncttrue
\mciteSetBstMidEndSepPunct{\mcitedefaultmidpunct}
{\mcitedefaultendpunct}{\mcitedefaultseppunct}\relax
\EndOfBibitem
\bibitem[Pettersen \emph{et~al.}(2004)Pettersen, Goddard, Huang, Couch,
  Greenblatt, Meng, and Ferrin]{Pettersen2004}
E.~F. Pettersen, T.~D. Goddard, C.~C. Huang, G.~S. Couch, D.~M. Greenblatt,
  E.~C. Meng and T.~E. Ferrin, \emph{J. Comput. Chem.}, 2004, \textbf{25},
  1605--1612\relax
\mciteBstWouldAddEndPuncttrue
\mciteSetBstMidEndSepPunct{\mcitedefaultmidpunct}
{\mcitedefaultendpunct}{\mcitedefaultseppunct}\relax
\EndOfBibitem
\bibitem[Gojzewski \emph{et~al.}(2012)Gojzewski, Makowski, Hashim, Kopcansky,
  Tomori, and Timko]{Gojzewski2012}
H.~Gojzewski, M.~Makowski, A.~Hashim, P.~Kopcansky, Z.~Tomori and M.~Timko,
  \emph{Scanning}, 2012, \textbf{34}, 159--169\relax
\mciteBstWouldAddEndPuncttrue
\mciteSetBstMidEndSepPunct{\mcitedefaultmidpunct}
{\mcitedefaultendpunct}{\mcitedefaultseppunct}\relax
\EndOfBibitem
\bibitem[Glatter(1977)]{glatter1977new}
O.~Glatter, \emph{J. Appl. Crystallogr.}, 1977, \textbf{10}, 415--421\relax
\mciteBstWouldAddEndPuncttrue
\mciteSetBstMidEndSepPunct{\mcitedefaultmidpunct}
{\mcitedefaultendpunct}{\mcitedefaultseppunct}\relax
\EndOfBibitem
\bibitem[Glatter(1979)]{glatter1979interpretation}
O.~Glatter, \emph{J. Appl. Crystallogr.}, 1979, \textbf{12}, 166--175\relax
\mciteBstWouldAddEndPuncttrue
\mciteSetBstMidEndSepPunct{\mcitedefaultmidpunct}
{\mcitedefaultendpunct}{\mcitedefaultseppunct}\relax
\EndOfBibitem
\bibitem[Vestergaard and Hansen(2006)]{Vestergaard:wf5022}
B.~Vestergaard and S.~Hansen, \emph{J. Appl. Crystallogr.}, 2006, \textbf{39},
  797--804\relax
\mciteBstWouldAddEndPuncttrue
\mciteSetBstMidEndSepPunct{\mcitedefaultmidpunct}
{\mcitedefaultendpunct}{\mcitedefaultseppunct}\relax
\EndOfBibitem
\bibitem[Bender \emph{et~al.}(2017)Bender, Bogart, Posth, Szczerba, Rogers,
  Castro, Nilsson, Zeng, Sugunan, Sommertune,\emph{et~al.}]{Bender2017}
P.~Bender, L.~Bogart, O.~Posth, W.~Szczerba, S.~Rogers, A.~Castro, L.~Nilsson,
  L.~Zeng, A.~Sugunan, J.~Sommertune \emph{et~al.}, \emph{Sci. Rep.}, 2017,
  \textbf{7}, 45990\relax
\mciteBstWouldAddEndPuncttrue
\mciteSetBstMidEndSepPunct{\mcitedefaultmidpunct}
{\mcitedefaultendpunct}{\mcitedefaultseppunct}\relax
\EndOfBibitem
\bibitem[Dewhurst \emph{et~al.}(2016)Dewhurst, Grillo, Honecker, Bonnaud,
  Jacques, Amrouni, Perillo-Marcone, Manzin, and Cubitt]{Dewhurst2016}
C.~D. Dewhurst, I.~Grillo, D.~Honecker, M.~Bonnaud, M.~Jacques, C.~Amrouni,
  A.~Perillo-Marcone, G.~Manzin and R.~Cubitt, \emph{J. Appl. Crystallogr.},
  2016, \textbf{49}, 1--14\relax
\mciteBstWouldAddEndPuncttrue
\mciteSetBstMidEndSepPunct{\mcitedefaultmidpunct}
{\mcitedefaultendpunct}{\mcitedefaultseppunct}\relax
\EndOfBibitem
\bibitem[Bender \emph{et~al.}(2017)Bender, Fern\'{a}ndez~Barqu\'{\i}n,
  Fdez-Gubieda, Gonz\'{a}lez-Alonso, Honecker, Marcano, and
  Szczerba]{Bender2017b}
P.~Bender, L.~Fern\'{a}ndez~Barqu\'{\i}n, M.~Fdez-Gubieda,
  D.~Gonz\'{a}lez-Alonso, D.~Honecker, L.~Marcano and W.~Szczerba, 2017\relax
\mciteBstWouldAddEndPuncttrue
\mciteSetBstMidEndSepPunct{\mcitedefaultmidpunct}
{\mcitedefaultendpunct}{\mcitedefaultseppunct}\relax
\EndOfBibitem
\bibitem[Honecker \emph{et~al.}(2010)Honecker, Ferdinand, D{\"o}brich,
  Dewhurst, Wiedenmann, G{\'o}mez-Polo, Suzuki, and Michels]{honecker2010}
D.~Honecker, A.~Ferdinand, F.~D{\"o}brich, C.~Dewhurst, A.~Wiedenmann,
  C.~G{\'o}mez-Polo, K.~Suzuki and A.~Michels, \emph{Eur. Phys. J. B}, 2010,
  \textbf{76}, 209--213\relax
\mciteBstWouldAddEndPuncttrue
\mciteSetBstMidEndSepPunct{\mcitedefaultmidpunct}
{\mcitedefaultendpunct}{\mcitedefaultseppunct}\relax
\EndOfBibitem
\end{mcitethebibliography}

\providecommand*{\mcitethebibliography}{\thebibliography}
\csname @ifundefined\endcsname{endmcitethebibliography}
{\let\endmcitethebibliography\endthebibliography}{}

\end{document}